Styles Needed: doublespace.sty, cite.sty, equations.sty
	       If you have access to FTP fetch them from
	       sun.soe.clarkson.edu in pub/tex/latex-styles
	       directory via anonymous ftp.
	       For those with e-mail access only one can request them
	       from  LISTSERV@DHDURZ1.BITNET or FILESERV@SHSU.BITNET
	       (send a small message with the line 'help' and
	       instructions will be sent to you)

\documentstyle[12pt,equations,doublespace,cite]{article}

\newcommand{\be}{\begin{equation}}
\newcommand{\ee}{\end{equation}}

\newcommand{\bea}{\begin{eqnarray}}
\newcommand{\eea}{\end{eqnarray}}

\newcommand{\bsub}{\begin{subequations} }
\newcommand{\esub}{\end{subequations} }

\newenvironment{appendletter} 
 {
  \setcounter{equation}{0} 
  
 }{
   \typeout{Appendix *\thesection*  done}
   \stepcounter{section} 
 }

\def\lequiv{\raise 0.4ex \hbox{$<$} \kern -0.8 em \lower 0.62 ex \hbox{$\sim$}
}
\def\gequiv{\raise 0.4ex \hbox{$>$} \kern -0.7 em \lower 0.62 ex \hbox{$\sim$}
}

\newcommand{\drI}{\mbox{$(\Delta r)_{I}$}}
\newcommand{\drII}{\mbox{$(\Delta r)_{II}$}}

\newcommand{\mwI}{\mbox{$(\mw)_{I}$}}
\newcommand{\mwII}{\mbox{$(\mw)_{II}$}}

\def\Im{\mathop{{\cal I}\!m}}
\def\Re{\mathop{{\cal R}\!e}}

\newcommand{\dr}{\mbox{$\Delta r$}}
\newcommand{\drcarw}{\mbox{$\Delta \hat{r}_{{\scriptscriptstyle W}}$}}
\newcommand{\drcar}{\mbox{$\Delta \hat{r}$}}

\newcommand{\ms}{\mbox{$\overline{\scriptstyle MS}$}}

\newcommand{\sms}{\mbox{$\overline{\scriptscriptstyle MS}$}}

\newcommand{\sinw}{\mbox{$\sin^{2}\!\theta_{\scriptscriptstyle W}$}}

\newcommand{\sincarmz}{\mbox{$\sin^{2}\!\hat{\theta}_{\scriptscriptstyle
W}(\mz)$}}
\newcommand{\coscarmz}{\mbox{$\cos^{2}\!\hat{\theta}_{\scriptscriptstyle
 W}(\mz)$}}

\newcommand{\MeV}{\mbox{~MeV}}
\newcommand{\GeV}{\mbox{~GeV}}
\newcommand{\TeV}{\mbox{~TeV}}

\newcommand{\Gmu}{\mbox{$G_\mu$}}

\newcommand{\mf}{\mbox{$m_{f}$}}

\newcommand{\mt}{\mbox{$m_{t}$}}
\newcommand{\mts}{\mbox{$m_{t}^{2}$}}

\newcommand{\mh}{\mbox{$m_{\scriptscriptstyle H}$}}

\newcommand{\mz}{\mbox{$m_{\scriptscriptstyle Z}$}}
\newcommand{\mzs}{\mbox{$m^{2}_{\scriptscriptstyle Z}$}}

\newcommand{\mw}{\mbox{$m_{\scriptscriptstyle W}$}}
\newcommand{\mws}{\mbox{$m^{2}_{\scriptscriptstyle W}$}}

\newcommand{\Z}{\mbox{$Z^{0}$}}
\newcommand{\Wpm}{\mbox{$W^{\pm}$}}

\newcommand{\scs}{\mbox{$\hat{s}^{2}$}}

\newcommand{\ccs}{\mbox{$\hat{c}^{2}$}}

\newcommand{\kcq}{\mbox{$\hat{\kappa}(q^2)$}}

\newcommand{\Agz}{\mbox{$A_{\scriptscriptstyle\gamma Z}$}}

\newcommand{\Azz}{\mbox{$A_{\scriptscriptstyle Z Z}$}}
\newcommand{\Azzp}[1]{\mbox{$A_{\scriptscriptstyle Z Z}^{#1}$}}

\newcommand{\Aww}{\mbox{$A_{\scriptscriptstyle W W}$}}
\newcommand{\Awwp}[1]{\mbox{$A_{\scriptscriptstyle W W}^{#1}$}}

\newcommand{\Ord}[1]{\mbox{${\cal O}\!\left(#1\right)$}}

\newcommand{\pigg}{\mbox{$\Pi_{\gamma\gamma}$}}

\newcommand{\piggp}[1]{\mbox{$\Pi^{\mbox{\scriptsize{(#1)}}}_{\gamma\gamma}$}}

\newcommand{\Drhob}{\mbox{$\Delta\bar{\rho}$}}

\newcommand{\rhoh}{\mbox{$\hat{\rho}$}}

\newcommand{\ehs}{\mbox{$\hat{e}^{2}$}}
\newcommand{\alphah}{\mbox{$\hat{\alpha}$}}

\newcommand{\sltdeoe}{\mbox{${\displaystyle 2\delta\!\mbox{\em
e}}/\!{\displaystyle \mbox{\em e}}$}}

\newcommand{\da}{\mbox{$\Delta\alpha$}}

\newcommand{\bra}[1]{\mbox{$\langle#1\mid$}}
\newcommand{\ket}[1]{\mbox{$\mid #1\rangle$}}

\renewcommand{\piggp}[1]{\mbox{$\Pi^{\mbox{\scriptsize{($#1$)}}}_{\gamma
\gamma}$}}

\begin{document}
\begin{titlepage}
\begin{flushright}
CERN-TH.6749/92\\
NYU-Th-92/12/05
\end{flushright}
\vspace{0.5 true cm}
\begin{center}
 {\Large
 Incorporation of QCD Effects in \\
 Basic Corrections of the Electroweak Theory  }\\[1cm]

 {\sc Sergio Fanchiotti},${}^{\ast}$ {\sc Bernd Kniehl},${}^{\dagger}$
 and {\sc Alberto Sirlin}\phantom{,}${}^{\ddagger}$\\[1cm]
 \begin{tabular}{l}
 ${}^\ast$~{Theory Division, CERN, 1211~Geneva~23, Switzerland}\\[0.3cm]
 ${}^\dagger$~II.~Institut f\"ur Theoretische Physik, Universit\"at Hamburg,\\
 \phantom{${}^\dagger$}~Luruper Chaussee~149, 2000~Hamburg~50, Germany\\[0.3cm]
 ${}^\ddagger$~Department of Physics, New York University,\\
 \phantom{${}^\ddagger$}~4 Washington Place, New York, NY~10003, USA
 \end{tabular}
\end{center}
\vspace{0.5cm}
\begin{abstract}
 We study the incorporation of QCD effects in the basic electroweak
 corrections \drcar, \drcarw, and \dr. They include perturbative
 \Ord{\alpha\alpha_s} contributions and $t\bar{t}$ threshold effects.
 The latter are studied in the resonance and Green-function approaches,
 in the framework of dispersion relations that automatically satisfy
 relevant Ward identities. Refinements in the treatment of the
 electroweak corrections, in both the \ms\ and the on-shell schemes of
 renormalization, are introduced, including the decoupling of the top
 quark in certain amplitudes, its effect on $\hat{e}^2(\mz)$ and
 \sincarmz, the incorporation of recent results on the leading irreducible
 \Ord{\alpha^2} corrections, and simple expressions for the residual,
 i.e.\  ``non-electromagnetic'', parts of \drcar, \drcarw, and \dr.
 The results are used to obtain accurate values for \mw\ and
 \sincarmz, as functions of \mt\ and \mh. The higher-order effects induce
 shifts in these parameters comparable to the expected experimental
 accuracy, and they increase the prediction for \mt\ derived from
 current measurements. The \ms\ and the on-shell calculations of \dr, in a
 recently proposed formulation, are compared and found to be in
 excellent agreement over the wide ranges $60\GeV \leq \mh \leq 1 \TeV$,
 $\mz \leq \mt \leq 250 \GeV$.
\end{abstract}
\vspace{0.2cm}\nopagebreak
{\small \noindent CERN-TH.6749/92\\
 December~1992}
\end{titlepage}

\section{Introduction}
It has been recently shown~\cite{R1} that the \ms\ method of
renormalization provides a very convenient framework to incorporate
higher-order corrections to \dr~\cite{R2} arising from resummation
of one-loop effects.
These include not only leading logarithms of \Ord{(\alpha \ln(\mz/\mf))^n},
where \mf\ is a generic fermion mass, and subleading
logarithms of \Ord{\alpha^2 \ln(\mz/\mf)}~\cite{R3},
but also terms of \Ord{(\alpha\,\mts/\mws)^n}~\cite{R1,R4,R5}.
The reason can be traced to the fact that in this method one essentially
subtracts the divergent parts of the amplitudes.
In contrast with other approaches, this procedure circumvents the
introduction of mass singularities and \Ord{\alpha\,\mts/\mws} terms via
the finite parts of the counterterms. As a consequence, the renormalized
perturbative expansion has a structure very similar to that of the bare
theory, where resummation of one-loop effects is easy to
implement~\cite{R1,R4}.
There are, of course, irreducible two-loop contributions of
\Ord{(\alpha\,\mts/\mws)^2}.
As discussed in Refs.\cite{R1,R4,R5}, these
can be gleaned from Ref.\cite{R6} and the more recent work of
Ref.\cite{R7} on the \Ord{\alpha^2} corrections to the
$\rho$ parameter.

It has also been recently shown~\cite{R8} that it is possible to
derive a simple and accurate expression for \dr, within the on-shell
method of renormalization~\cite{R2}, which contains the same leading and
subleading contributions described above.
On the other hand, the irreducible two-loop contributions of
\Ord{\alpha^2 \mts/\mws} have not been computed, so that both the \ms\ and
on-shell calculations of \dr\ become uncertain at this level of accuracy.
In fact, one can see that the difference between the two calculations
and their inherent theoretical uncertainty due to the neglect of
higher-order electroweak corrections start with subleading terms of
\Ord{(\alpha/\pi s^2)(c^2/s^2) x_t}, where $x_t$ is the leading correction
to the $\rho$ parameter (cf.\ Eq.~(\ref{E17b})),
or, equivalently, of \Ord{(c^2/s^2)(\alpha_2/2\pi)(\kappa_t/2\pi)}, where
$\alpha_2 \equiv g^2/4\pi$ and $\kappa_t \equiv (G_t)^2/4\pi$ are the
SU(2) and the Yukawa couplings of the top quark, respectively.
These are very small for low \mt\ and are expected to be $\approx
8\times10^{-4}$ for $\mt = 250 \GeV$.
(As we will see, over the large ranges $\mz\le\mt\le250\GeV$,
$60\GeV\le\mh\le1\TeV$, the actual numerical evaluation of the on-shell
expression of Ref.\cite{R8} and the \ms\ calculations show a very small
difference, reaching a maximum value of $2.5\times10^{-4}$ at $\mt=250\GeV$
and $\mh=1\TeV$, a very precise agreement which may be somewhat fortuitous.)
This uncertainty is to be compared with an estimated error of
$\pm 9\times10^{-4}$ originating in the \Ord{\alpha} contributions of the
first five quark flavors~\cite{R9,R10}.

In order to  set the stage for our discussion, it is convenient
to recall at this point some of the basic relations of the on-shell  and
\ms\ frameworks~\cite{R1,R2,R3,R11,R12}:
\bea
       s^2 &=& \frac{A^2}{\mws (1-\dr)},          \label{E1}   \\*
       \scs &=& \frac{A^2}{\mws (1-\drcarw)},     \label{E2}   \\*
       \scs \ccs &=& \frac{A^2}{\mzs (1-\drcar)}, \label{E3}
\eea
where \mw\ and \mz\ are the physical masses,
$A =\left(\pi\alpha/(\sqrt{2}\Gmu)\right)^{1/2} = (37.2802\pm 0.0003) \GeV$,
$s^2$, \scs, and \ccs\ are abbreviations for $\sinw \equiv 1-\mws/\mzs$
and the \ms\ parameters
\sincarmz\ and \coscarmz, respectively,
and \dr, \drcarw, and \drcar\ are radiative corrections.
The 't~Hooft mass, $\mu$, has been set equal to \mz\ in
Eqs.~(\ref{E2}) and (\ref{E3}).

It follows from the analysis of Ref.\cite{R13} that \mz\ and \mw\ in
Eqs.~(\ref{E1}) and (\ref{E2}) can be identified, phenomenologically, with the
masses measured in current experiments and, theoretically, with the
definition $m_1^2 = m_2^2 + \Gamma_2^2$, where $\bar{s} = m_2^2 - i\,m_2
\Gamma_2$ is the relevant complex pole position~\cite{R14}.
The latter is given by $\bar{s} = m_0^2 + A(\bar{s})$, where $m_0$ is
the bare mass and $A(s)$ the conventional self energy, which includes
tadpoles, tadpole counterterms, and, in the \Z\ case, $\gamma Z$ mixing
effects that start in \Ord{\alpha^2}.
On general grounds, it is expected that $m_2$ and $\Gamma_2$, and
therefore $m_1$, are gauge-invariant to all
orders~\cite{R13,R15}.
Over a large class of gauges, including those in which explicit
calculations have been carried out, $m_1$ differs from the
``field-theoretic'' or ``on-shell'' definition, $m^2 = m_0^2 + \Re A(m^2)$,
by gauge-dependent terms of \Ord{\alpha^3}~\cite{R13}.
Because contributions of this order are well beyond the accuracy that
may be achieved in the foreseeable future, the replacement of $m$ by
the more rigorous definition $m_1$ does not require a modification of
the radiative corrections \dr, \drcarw, and \drcar.
Using the expression
\be
        s^2 = \frac{1}{2} \left\{
	   1 - \left[ 1 - \frac{4 A^2}{\mzs (1-\dr)} \right]^{1/2}
	\right\} ,
\label{E4}
\ee
equivalent to Eq.~(\ref{E1}), the analogous one with $s^2 \rightarrow
\scs$ and $\dr \rightarrow \drcar$, equivalent to Eq.~(\ref{E3}), and the
accurately known value $\mz = (91.187 \pm 0.007) \GeV$~\cite{R16},
the corrections \dr\ and \drcar\ lead to precise evaluations of $s^2$
and \scs, as functions of \mt\ and \mh.
These, in turn, can be compared with other determinations of $s^2$ and
\scs\ to constrain the value of \mt\ and, in the future, that of \mh.
They are also important input parameters in the prediction of the \Z\
partial widths and on-resonance asymmetries, as some of these
observables depend very sensitively on the weak mixing angle (see, for
example, Ref.\cite{R17}).
More generally, the basic corrections \dr, \drcarw, and \drcar\ are
frequently employed to verify the Standard Model (SM) at the level of
its quantum corrections and in searches for signals of new
physics~\cite{R18,R19,R20,R21}.
It was also explained in Ref.\cite{R1} how \drcar\ and
\drcarw, relevant corrections in the \ms\ framework, can be employed to
evaluate the on-shell quantity \dr.

The aim of the present paper is to incorporate the leading QCD
corrections in the calculations of the radiative corrections \dr,
\drcar, and \drcarw. We also introduce some refinements in our previous
treatment of the electroweak corrections~\cite{R1}.

The relevant QCD contributions occur in the vacuum-polarization
functions associated with the \Wpm\ and \Z\ bosons and have been
extensively discussed in the
literature~\cite{R22,R23,R24,R25,R26,R27,R28,R29,R30,R31}.
In particular, the QCD corrections involving the $(t,b)$ isodoublet are
known to be significant for large \mt. There are actually two types of
effects that may be classified as perturbative \Ord{\alpha \alpha_s}
and threshold contributions.
In the literature, the latter are frequently referred to as
``non-perturbative.''
The perturbative \Ord{\alpha \alpha_s} components have been studied with
two different methods:
{\it i}) direct evaluation of the two-loop Feynman diagrams in
dimensional regularization, an approach that goes back to the pioneering
work of Djouadi and Verzegnassi~\cite{R22};
{\it ii}) calculation of the imaginary parts and computation of the full
amplitude by means of suitably defined dispersion
relations~(DRs)~\cite{R23,R24,R25,R26}.
It has been shown~\cite{R30,R31} that the two approaches are equivalent
in the evaluation of the perturbative contributions to \dr\ and
$\Delta\rho$, a welcome fact.
On the other hand, the \ms\ scheme is implemented in the framework of
dimensional regularization and, for a full determination of the
subtraction constants, one must appeal to method ({\it i}).

Threshold effects on the absorptive parts of the self energies have, in
turn, been discussed in two different approaches:
{\it a}) in Ref.\cite{R24} the contributions of densely spaced, narrow
quarkonium resonances were taken into account on the basis of a specific
quark-antiquark potential;
{\it b}) in Refs.\cite{R27,R28} one considers the imaginary part of the Green
function for the non-relativistic Schr\"odinger equation that
characterizes the $t\bar{t}$ system near threshold.
The latter formulation effectively resums the contributions of soft-gluon
exchanges in the ladder approximation (see also Ref.\cite{R29}).

For sufficiently low \mt\ there should be, near threshold, a rich
spectrum of distinct non-relativistic states bound by strong long-range
forces and the approach ({\it a}) is very natural.
For increasing \mt, however, the weak decay of a single top quark inside
the bound states becomes important and, for $\mt\ \gequiv\ 130\GeV$, the
partial width of $t \rightarrow W^+\,b$ is so large that the revolution
period of a $t \bar{t}$ bound state would exceed its lifetime.
As a consequence, the individual resonances lose their distinctiveness
and are smeared out to a coherent structure~\cite{R27,R28,R29}.
In that regime, the Green-function method is more appropriate.
In summary, one expects approaches ({\it a}) and ({\it b}) to be
preferable for lower ($ \lequiv 130 \GeV$) and higher
($ \gequiv 130 \GeV$) values of \mt.
Both formulations deal directly with the absorptive parts of the amplitudes.
To obtain the real parts is then necessary to employ DRs.
This was done in detail in Ref.\cite{R31} using DRs for the
vacuum-polarization functions directly constructed from relevant Ward
identities~\cite{R30}.
In conjunction with very plausible assumptions concerning the asymptotic
behavior  of the threshold effects for large $q^2$, this procedure
leads to specific results for the real parts.

In Section~7, we compare four different calculations of \mw\ and
\sincarmz, as functions of \mt\ and \mh:
{\it i}) only electroweak corrections;
{\it ii}) electroweak plus perturbative \Ord{\alpha \alpha_s} corrections;
{\it iii}) the above, plus threshold effects in the resonance approach;
{\it iv}) same as ({\it iii}) with threshold effects in the Green-function
framework.
This allows us to demonstrate the magnitude of the QCD corrections and,
at the same time, to separate the threshold  effects from the more established
perturbative \Ord{\alpha \alpha_s} contributions.
Although the two approaches to treat the threshold effects are
certainly not identical, we find the encouraging and fortunate result
that their overall effects on \mw\ and \sincarmz\ are quite close over the
entire range $\mz \leq \mt \leq 250 \GeV$.

As mentioned before, aside from incorporating the QCD effects in the
relevant self energies, we introduce some refinements in our treatment
of the \Ord{\alpha} electroweak corrections.
Conceptually, the most interesting modification is a slight change in
the definition of the fundamental couplings $\ehs(\mz)$ and \sincarmz\
of the \ms\ scheme, which is introduced in order to make them essentially
independent of
heavy particles such as the top quark or unknown massive excitations.
In the case of \sincarmz, we follow a convention recently proposed at the
one-loop level by Marciano and Rosner~\cite{R32,R33}, and explain how to
extend it when \Ord{\alpha \alpha_s} corrections are included.
We emphasize that these modifications in the definitions of the
fundamental \ms\ parameters do not affect, to the order of the
calculations, the relations between physical observables because they
are compensated by corresponding changes in the appropriate radiative
corrections.
A second change is that we use an updated calculation by
Jegerlehner~\cite{R10} for the contribution of the first five quark
flavors to the photon vacuum-polarization function.
A third modification is that we incorporate the very recent results of
Ref.\cite{R7} concerning the irreducible two-loop corrections of
\Ord{(\alpha\,\mts/\mws)^2}.
In the rest of the calculations, as we did in Ref.\cite{R1}, we
treat the $u$, $d$, and $s$ quarks as massless but we  now include terms of
\Ord{\alpha\,m_f^2/\mws}, where $f = c, b, \tau,\ldots$\ .
Although they are very small---they contribute to \dr\ only at the
$\lequiv\,1\times10^{-4}$ level---, their incorporation may facilitate detailed
comparisons with calculations by other authors.

The plan of the paper is the following:
in Section~2, we discuss the definitions of the basic \ms\ parameters
$\ehs(\mz)$ and \sincarmz\ in the presence of the \Ord{\alpha \alpha_s}
corrections, with emphasis on the decoupling of heavy particles.
In Section~3, we incorporate the perturbative \Ord{\alpha \alpha_s}
contributions in \drcarw\ and \drcar.
We emphasize the important fact that the magnitude of the \Ord{\alpha \alpha_s}
effects depends sensitively on the precise definition of \mt.
Our detailed calculations, as well as the other papers in the
literature, employ the ``on-shell'' definition of \mt.
In the discussion we give a brief argument to indicate why this choice
is useful and appropriate.
In Section~4, we present a simple method to separate the residual,
i.e.\  ``non-electromagnetic'', parts of \drcar, \drcarw, and \dr.
In Section~5, we include the perturbative \Ord{\alpha \alpha_s}
corrections in the calculation of \dr\ in the on-shell scheme, using the
formulation of Ref.\cite{R8}.
In Section~6, we discuss the incorporation of threshold effects in
\drcar, \drcarw, and \dr.
We include estimates, based on a simple ``Bohr-atom'' model, of the
contribution of the 1S toponium resonance to the imaginary part of the
self energies.
We find that this simple model gives values roughly similar to the
calculations carried out with more realistic quark--antiquark potentials.
In Section~7, we use the theoretical results to carry out precise
calculations of \mw\ and \sincarmz, in the manner explained before.
We find that the QCD and other higher-order corrections induce shifts in
\mw\ and \sincarmz\ comparable with the expected experimental precision.
Interestingly, all of them increase the value of \mt\ derived from
current measurements.
We also compare the \ms\ and on-shell calculation of \dr, in
the formulation of Ref.\cite{R8}.
We find that, as was the case in the absence of perturbative
\Ord{\alpha \alpha_s} corrections, the \ms\ and on-shell calculations of
\dr\ are very close over a large range of \mt\ and \mh\ values.
The Appendices discuss basic expressions for the perturbative
\Ord{\alpha \alpha_s} corrections, the effect of top-quark decoupling in
\drcarw\ and \drcar, and the very small contribution from finite fermion
masses.
\section{The Parameters $\ehs(\mz)$ and \sincarmz}

In our previous treatment~\cite{R1} we defined these
parameters, at the one-loop level, by simply subtracting from the
radiatively corrected cofactors the terms involving
\bsub
\be
	\delta = {1\over n-4} + {1\over2}[\gamma-\ln(4\pi)] ,
\label{E5a}
\ee
and setting the 't~Hooft mass scale, $\mu$, equal to \mz.
Because at one loop $\delta$ always occurs in combination with
$-\ln\mu$, this is equivalent to subtracting only the pole terms,
$(n-4)^{-1}$, rescaling $\mu$ according to
\be
	\mu = {\mu'\,e^{\gamma/2}\over (4\pi)^{1/2}},
\label{E5b}
\ee
\esub
and then setting $\mu' = \mz$. The second formulation can be
conveniently generalized to higher-order corrections and one can define
the \ms\ renormalization procedure as the subtraction of pole terms of
the form $(n-4)^{-m}$, where $m$ is an integer $\geq 1$, and the identification
of the rescaled 't~Hooft parameter $\mu'$ with the relevant mass scale,
in this case \mz.
As is well known, the factor $e^{\gamma/2}(4\pi)^{-1/2}$ is appended in
Eq.~(\ref{E5b}) to cancel relatively large numerical constants that are
an artifact of dimensional regularization~\cite{R34}.

In Ref.\cite{R1} we applied this procedure uniformly,
independently of whether the top quark is more or less massive than \mz.
On the other hand, it is desirable to treat heavy
particles, as much as possible, as decoupled. For example, when $\mt >\mz$
it is convenient to subtract from the amplitude terms involving
$\ln(\mt/\mz)$ and to absorb them in the coupling constants.
In Ref.\cite{R1} we did not follow this route for two reasons:
{\it i}) \sincarmz\ appears as a cofactor in several important
radiatively corrected amplitudes and it is not possible to absorb
completely the $\ln(\mt/\mz)$ terms occurring in all of them;
{\it ii}) some important relations, such as Eq.~(\ref{E3}), contain terms
proportional to \mts, which certainly do not decouple.
Recently, however, Marciano and Rosner~\cite{R32,R33} proposed to
implement the decoupling idea, at the one-loop level, by absorbing in
\sincarmz\ all $\ln(m/\mz)$ terms with $m > \mz$ occurring in a specific
amplitude, namely the $\gamma Z$ self energy evaluated at $q^2 = \mzs$,
$\Re\Agz(\mzs)$.
Here $m$ is the mass of the top quark or any unknown heavy particle with
$m > \mz$.
With this convention, a heavy top or a heavy unknown particle
decouples in the limit $m/\mz \gg 1$ from the cofactor \kcq\
multiplying \scs\ in most neutral-current processes; as a
consequence, this parameter can effectively be determined from the
on-resonance asymmetries without hindrance from unknown ``heavy
physics.''

We now explain how we implement the decoupling idea in the presence of
the \Ord{\alpha \alpha_s} corrections.
It is convenient to start with $\ehs(\mz)$, which we frequently
abbreviate as \ehs.
We recall the relation between the bare charge $e_0$ and the
conventional renormalized charge~$e$:
\be
    e^2 = e^2_0 \left\{
    		1 + e^2_0 \piggp{f}(0) + \frac{7e^2_0}{8\pi^2}
	\left[
	 \frac{1}{n-4} + \ln\frac{\mw}{\mu'} - \frac{1}{21}
	\right]
    		\right\}^{-1} ,
\label{E6}
\ee
where $\piggp{f}(0)$ is the usual fermionic vacuum-polarization function
evaluated at $q^2 = 0$ and the last term represents \Ord{e^2_0} bosonic
contributions to charge renormalization that must be included in the SM.
The latter can be gleaned, for example, from Ref.\cite{R2}.
As explained in Refs.\cite{R1,R2}, because of the existence of
mass singularities associated with the light quarks, it is not possible
to calculate perturbatively the contribution of the first five quark
flavors to $\piggp{f}(0)$.
Calling this contribution $\piggp{5}(0)$, the problem is circumvented by
writing $\piggp{5}(0) = \Re\left(\piggp{5}(0) - \piggp{5}(\mzs)\right) +
\Re\piggp{5}(\mzs)$.
The first term represents the five-flavor contribution to the
renormalized photon vacuum-polarization function at $q^2 = \mzs$ and can be
evaluated using dispersion relations, experimental data on $e^+ e^-
\rightarrow hadrons$ and QCD corrections~\cite{R2,R9,R10}.
The  presence of a large invariant momentum, $q^2 = \mzs$, in the third
term prevents the occurrence of fermionic mass singularities and, as a
consequence, $\Re\piggp{5}(\mzs)$ can be analysed perturbatively.
The same is, of course, true for the leptonic and bosonic contributions
and, because of its large mass, for the top contribution $\piggp{t}(0)$.
Including irreducible two-loop contributions of \Ord{\alphah \alphah_s,
\alphah^2} to $\ehs \piggp{f}(0)$ arising from virtual gluon and photon
interchanges, we obtain (see Appendix A)
\bsub
\bea
   \ehs \piggp{f}(0) &=&
     \frac{2\alphah}{3\pi} \sum_l \left[ \ln{\mu'\over m_l}
     \left(1 + \frac{3 \alphah}{4 \pi}\right) +
     \frac{45\alphah}{32\pi} \right] \nonumber\\*
  & &  + \frac{8\alphah}{9\pi} \left[ \ln{\mu'\over\mt}
    \left(1 + \frac{\alphah_s}{\pi} + \frac{\alphah}{3 \pi}\right) +
    \frac{15}{8} \left(\frac{\alphah_s}{\pi} + \frac{\alphah}{3 \pi}\right)
    \right] \nonumber\\*
  & &  + \frac{\alphah}{\pi} \sum_{q \neq t} Q^2_q \left[
    2 \ln{\mu'\over\mz} \left(1 +
        \frac{\alphah_s}{\pi} + \frac{3 \alphah}{4 \pi} Q^2_q \right)
	+ f_1(r_q) + \left(\frac{\alphah_s}{\pi} + \frac{3 \alphah}{4 \pi}
	Q^2_q \right) f_2(r_q) \right] \nonumber\\*
  & &  + \ehs Re \left(\piggp{5}(0) - \piggp{5}(\mzs)\right)
    - \frac{\alphah}{\pi}\,\frac{I}{(n-4)} ,
\label{E7a}
\eea
where the $l$ and $q$ sums run over leptons and quarks, respectively,
the color factor 3 is henceforth explicitly included, $r_q = \mzs/\left(4
m_q^2\right)$,
\bea
	f_1(r) &=& \ln(4 r) - \left(2 + \frac{1}{r}\right)
	\left(1 - \frac{1}{r}\right)^{1/2}
	\cosh^{-1}\sqrt r + \frac{5}{3} + \frac{1}{r} \nonumber\\*
	&=& \frac{5}{3} + \frac{3}{2 r}
	+ \Ord{\frac{\displaystyle \ln r}{\displaystyle r^2}} , \\*
	f_2(r) &=& \ln(4 r) + {\Re V_1(r)\over r} - 4 \zeta(3) + \frac{55}{12} ,
	\label{E7b} \nonumber\\*
	&=& -4 \zeta(3) + \frac{55}{12} - \frac{3}{r} \ln(4 r)
	+ \Ord{\frac{\displaystyle\ln^2 r}{\displaystyle r^2}}
	\label{E7c} ,
\eea
$\zeta(3) = 1.20206\ldots$, $V_1(r)$ is a complicated function defined
in Ref.\cite{R25}, and
$I=16/3+5\alphah_s/(3\pi)+11\alphah/(9\pi)$.
The first and second terms in Eq.~(\ref{E7a}) are the finite parts of the
leptonic and top contributions, while the sum over $q$ is the finite part
of the perturbative evaluation of $\ehs\Re\piggp{5}(\mzs)$.
In the latter the terms proportional to $r_q^{-1}$ are extremely small
and we can replace the functions $f_1(r)$ and $f_2(r)$ by  their
asymptotic values $f_1(\infty) = 5/3$ and $f_2(\infty) = 55/12 - 4
\zeta(3) \approx -0.22491$.
The last term in Eq.~(\ref{E7a}) represents the divergent part up to
terms of \Ord{\alphah^2, \alphah \alphah_s}.
As is well known, up to two loops $\pigg(q^2)$ is linear in $(n-4)^{-1}$ and
$\ln \mu'$.
We note that the cofactors of $(n-4)^{-1}$ are equal to those of
$\ln(1/\mu')$ at one loop and to one-half those of $\ln(1/\mu')$ at
two loops.

In order to obtain the relation between \ehs\ and $e^2$, one
writes $e_0^2 = \ehs/\hat{Z}_{e}$ in Eq.~(\ref{E6}), uses the
counterterms present in $\hat{Z}_{e}$ to cancel the $(n-4)^{-1}$ terms
in Eqs.~(\ref{E6}) and (\ref{E7a}), and sets $\mu' = \mz$ in those
equations. The mass scale employed in $\alphah_s$ is discussed in
greater detail at the end of this section.
With the exception of the top quark, all the particles contributing to
Eqs.~(\ref{E6},\ref{E7a}) are less massive than \mz\ and their
contribution is retained.
To implement decoupling in Eq.~(\ref{E6}), we also subtract the finite
top contribution when $\mt > \mz$, so that $\hat{Z}_e$ in that case
contains an additional finite counterterm and reads
\be
  \hat{Z}_e = 1
 +{\alphah\over\pi}\left(I-{7\over2}\right){1\over n-4}
 - \left.\ehs\piggp{t}(0)\right|_{\sms}
 \theta(\mt-\mz) ,
 \label{E7d}
\ee
where \ms\ denotes the ``finite part'' after the \ms\ renormalization
has been carried out, i.e.\  the remainder after the $(n-4)$ poles
have been subtracted and $\mu'$ has been set equal to \mz, and the
superscript $(t)$ refers to the top-quark contribution.
Specifically,
\be
 \left.\ehs\piggp{t}(0)\right|_{\sms} =
  \frac{8\alphah}{9\pi}
   \left[ \ln{\mz\over\mt} \left(1 + \frac{\alphah_s}{\pi}
   + \frac{\alphah}{3\pi}\right)
     + \frac{15}{8} \left(\frac{\alphah_s}{\pi} + \frac{\alphah}{3\pi}\right)
   \right] .
\label{E7e}
\ee
\esub
The term proportional to $(n-4)^{-1}$ in Eq.~(\ref{E7d}) cancels the
divergent parts in Eqs.~(\ref{E6},\ref{E7a}).
The other term subtracts the finite top contribution, i.e.\  the
second term in Eq.~(\ref{E7a}), when $\mt > \mz$.
Because this contribution does not exactly vanish at $\mt = \mz$, the
above prescription leads to a small discontinuity at $\mt = \mz$.
An alternative that would ensure continuity would be to do the matching
at the point where the finite top correction vanishes, which is
$\mt =1.073\,\mz = 97.8\GeV$.
However, because the decoupling at $\mt > \mz$
is easy to implement and is analogous to what is done in some QCD
calculations~\cite{R35}, we will adopt it as our convention.

As the mass range $\mt<91\GeV$ has been excluded at the 95\% confidence
level \cite{R36}, we will
henceforth assume that $\mt > \mz$, in which case
Eqs.~(\ref{E6},\ref{E7a},d) lead to
\bsub
\be
  e^2 = {\ehs\over1 + (\alphah/\pi) \Delta_\gamma} ,
\label{E8a}
\ee
where
\bea
  \Delta_\gamma &=& \frac{7}{4} \ln{c^2} - \frac{1}{6}
     + \frac{2}{3}\sum_l \left[
       \ln{\mz\over m_l} \left(1 + \frac{3 \alphah(\mz)}{4 \pi}\right)
          + \frac{45\alphah(\mz)}{32\pi} \right]
	  \nonumber\\*
    & & + \frac{55}{27}
        + \left(\frac{11\alphah_s(\mz)}{9\pi}
 + \frac{35\alphah(\mz)}{108\pi}\right)
   \left(\frac{55}{12} - 4 \zeta(3) \right)
	  \nonumber\\*
    & & + 4 \pi^2 \Re \left(\piggp{5}(0) - \piggp{5}(\mzs)\right) ,
\label{E8b}
\eea
and $c^2$ is an abbreviation for
$\cos^2\theta_{\scriptscriptstyle W} \equiv \mws/\mzs$.
Solving for \ehs, we have
\be
  \ehs = {e^2\over1 - (\alpha/\pi) \Delta_\gamma} .
\label{E8c}
\ee
\esub
In $(\alpha/\pi) \Delta_{\gamma}$ we have retained very small terms of
\Ord{\alpha^2} arising from virtual-photon interchange. We have done
this because of the analogy with gluon contributions and the fact that
they contain interesting leptonic mass singularities.
It should be understood, however, that this does not represent a
complete \Ord{\alpha^2} calculation as there are other irreducible
two-loop contributions of this order arising from the bosonic sector and
from \Wpm, \Z, and $H$ interchanges in the fermionic sector.

Equation~(\ref{E8c}) allows us to compute \ehs\ in terms of $\alpha$,
independently of \mt\ or unknown particles heavier than \mz.
Using $\mz = 91.187\GeV$~\cite{R16}, $\mw = 80.22\GeV$~\cite{R37,R50},
$\alphah_s =\alphah_s(\mz) = 0.118$~\cite{R38}, and
$e^2\Re \left(\piggp{5}(0) - \piggp{5}(\mzs)\right) = 0.0282 \pm
0.0009$~\cite{R10}, we find $(\alpha/\pi) \Delta_{\gamma} = 0.0668 \pm 0.0009$
or $\alphah^{-1} = (4 \pi/\ehs) = 127.9 \pm 0.1$.
The expression for $(\alpha/\pi) \Delta_{\gamma}$ differs from the
quantity $- \sltdeoe|_{\sms}$ in Ref.\cite{R1} by the exclusion
of the $\ln(\mt/\mz)$ terms, a more accurate description of the QCD
corrections, and the updated calculation of the five-flavor contribution.
Numerically, however, $\alphah^{-1}$ is very close to the value $127.8
\pm 0.1$ reported in Ref.\cite{R1} for $\mt = \mz$, the small difference
essentially arising from the change in the five-flavor
contribution~\cite{R10}.

Concerning $\scs\equiv \sincarmz$, we recall that in the neutral-current
amplitudes this parameter is multiplied by the electroweak form factor
$\hat{\kappa}$, which contains the $\gamma Z$ mixing term
$-(\hat c/\hat s)\Agz(q^2)/q^2$~\cite{R17,R39}.
Here $\Agz(q^2)$ is the unrenormalized $\gamma Z$ transverse mixing
amplitude as defined in Ref.\cite{R40}, expressed in terms of the \ms\
couplings $\hat e$, $\hat s$, and $\hat c$.
In order to implement the decoupling, we apply the Marciano--Rosner
convention~\cite{R32,R33}, according to which the $\ln(\mt/\mz)$ terms in
$\Re \Agz(\mzs)/\mzs$ are subtracted for $\mt > \mz$.
At the two-loop level there is also an \mt-independent term, which
must be subtracted, too.
More generally, the idea is to subtract all contributions to $\Re
\Agz(\mzs)/\mzs$ that involve particles of mass $m > \mz$ and
do not decouple in the limit $m \rightarrow \infty$.
This can be implemented by adding a finite counterterm in the \scs\
renormalization, in analogy with Eq.~(\ref{E7d}).
Up to terms of \Ord{\alpha \alpha_s} we find,
in the case of the top quark,
\bsub
\be
	s_0^2 = \scs \left\{
		1 + {{\rm cte}\over n-4} +
 {\hat c\over\hat s}
 \left.\left(A_{\gamma Z}^{(t)}\right)'(0)\right|_{\sms}\right\},
\label{E9a}
\ee
where $s_0^2$ is the bare parameter,
$\left(A_{\gamma Z}^{(t)}\right)'(0)
=\left.(d/dq^2)A_{\gamma Z}^{(t)}(q^2)\right|_{q^2=0}$,
and \ms\ and the superscript $(t)$
have the same meaning as in Eq.~(\ref{E7d}).
Again, ${\rm cte}/(n-4)$ is the divergent part of the counterterm and the
last term is a finite contribution necessary to implement the decoupling
of the $\ln(\mt/\mz)$ and constant terms in the top contribution to
$\Re\Agz(\mzs)/\mzs$.
Specifically, we find:
\be
 {\hat c\over\hat s}
 \left.\left(A_{\gamma Z}^{(t)}\right)'(0)\right|_{\sms}=
 {\alphah\over\pi}\,d,
\label{E9b}
\ee
where
\be
	d = {1\over3}\left(\frac{1}{\scs} - \frac{8}{3}\right)
	    \left[ \ln{\mt\over\mz}\left(1 + \frac{\alphah_s}{\pi}\right)
	- \frac{15\alphah_s}{8 \pi}\right] .
\label{E9c}
\ee
\esub
In Eqs.~(\ref{E9a}--c) we have neglected all terms of \Ord{\alpha^2}.

We must still discuss the mass scale employed in $\alphah_s$.
Following Refs.\cite{R22,R24,R25,R26,R31},
in the present paper corrections arising
from the $(t,b)$ isodoublet are computed with $\alphah_s(\mt)$.
The reason is that the dominant contributions due to this isodoublet
involve mass scales of \Ord{\mt}~\cite{R31,R41}.
This choice can also be justified by arguments based on effective field
theory~\cite{R42}.
On the other hand, the perturbative contributions from the two
light-quark isodoublets involve self energies evaluated at $q^2 = \mws$ or
$q^2 = \mzs$ and, for that reason, are calculated with $\alphah(\mz)$.
As an example, in Eqs.~(\ref{E7e},\ref{E9c}), which involve top-quark
contributions, we identify $\alphah_s$ with $\alphah_s(\mt)$, while in
the perturbative part of Eq.~(\ref{E8b}), which includes light-quark
isodoublets, we employ $\alphah_s(\mz)$. (In the latter equation we make
a slight and negligible change to the above isodoublet rule by also
evaluating the very small bottom contribution with $\alphah_s(\mz)$.)
Numerically, the finite counterterms in Eqs.~(\ref{E7d},\ref{E9a}) are
quite small: for $\mt = 150 \GeV$, they are $1.0\times10^{-3}$ in
Eqs.~(\ref{E7d}--e) and $6.1\times10^{-4}$ in Eqs.~(\ref{E9b}--c);
for $\mt = 250 \GeV$, the corresponding values are $2.2\times10^{-3}$ and
$1.5\times10^{-3}$, respectively.
As we will see, when \scs\ is defined according to the decoupling
convention explained above, the finite counterterm in
Eq.~(\ref{E9a}) introduces small compensatory shifts in \drcarw\ and \drcar.
Similarly, $\hat{Z}_e$, defined in Eq.~(\ref{E7d}), will introduce small
compensatory changes in radiative corrections whenever \ehs\ is employed
as zeroth-order parameter.

\section{Perturbative \Ord{\alpha \alpha_s} Corrections to
         \drcarw, \drcar, and \dr }

When the decoupling of the top quark is implemented according to the
discussion of Section~2, the expression for \drcarw\ (cf.\ Eq.(\ref{E2}))
becomes (see Appendix B)
\bsub
\bea
   \drcarw &=& \left[ \frac{e^2}{\scs} \Re\left(\frac{\Aww(\mws)
                - \Aww(0)}{\mws} \right) + e^2  \piggp{f}(0)
             \right]_{\sms}
	     + \frac{\alpha}{\pi} \left(\frac{7}{4}\ln{c^2} - \frac{1}{6}
	                          \right) \nonumber\\*
  & & +\frac{e^2}{16\pi^2 \scs} \left\{
         6 + \frac{\ln{c^2}}{s^2} \left[
  \frac{7}{2} - \frac{5}{2}s^2 - \scs \left(5 - \frac{3 c^2}{2 \ccs}\right)
                                  \right] \right\}
  - \frac{\alpha}{\pi}\,d,
\label{E10a}
\eea
where $\Aww(q^2)$ is the unrenormalized $WW$ transverse self energy with
the coupling $e^2/\scs$ factored out~\cite{R1}, $d$ is defined in
Eq.~(\ref{E9c}), and the \ms\ symbol has the same meaning as in Section~2.
The quantities \drcarw, \drcar, and \dr\ are gauge-invariant but some
of their partial bosonic contributions are not.
In Eq.~(\ref{E10a}) and henceforth all explicit expressions and partial
contributions are given in the 't~Hooft--Feynman gauge.
In particular, the expression involving curly brackets represents vertex-
and box-diagram corrections to $\mu$ decay, evaluated in that gauge.

Except for the last term, Eq.~(\ref{E10a}) is the same as Eq.~(8b) in
Ref.\cite{R1}.
As \scs\ enters Eq.~(\ref{E2}) as a zeroth-order parameter and is
defined in the present paper according to the decoupling convention
explained in Section~2, there is now an additional contribution
$-(\alpha/\pi) d$ arising from the finite contribution in Eqs.~(\ref{E9a}--c).

We now discuss the evaluation of Eq.~(\ref{E10a}). The quantity
$\left[e^2 \piggp{f}(0)\right]_{\sms}$ can be obtained from Eq.~(\ref{E7a}).
We have
\bea
  \left[e^2 \piggp{f}(0)\right]_{\sms} + \frac{\alpha}{\pi} \left( \frac{7}{4}
  \ln{c^2} - \frac{1}{6}\right) &=&
  \frac{\alpha}{\pi} \left\{ \Delta_{\gamma}
  + \frac{8}{9} \left[ \ln\frac{\mz}{\mt}
  \left(1+\frac{\alphah_s}{\pi}
  + \frac{\alphah}{3 \pi} \right) \right.\right.
  \nonumber \\*
  & & \left.\left.
  + \frac{15}{8} \left(\frac{\alphah_s}{\pi}
    + \frac{\alphah}{3 \pi}\right) \right]\right\} ,
\label{E10b}
\eea
where $\Delta_{\gamma}$ is defined in Eq.~(\ref{E8b}) and the other terms
on the r.h.s.\ represent the top contribution.
Inserting Eq.~(\ref{E10b})  into Eq.~(\ref{E10a}) and neglecting small terms
of \Ord{\alpha^2}, we obtain
\bea
 \drcarw &=& \left[ \frac{e^2}{\scs}\Re\left(\frac{\Aww(\mws) - \Aww(0)}{\mws}
                                        \right) \right]_{\sms}
	   + \frac{\alpha}{\pi} \Delta_{\gamma} \nonumber\\*
& & +\frac{e^2}{16 \pi^2 \scs} \left\{6 +
 \frac{\ln{c^2}}{s^2} \left[
   \frac{7}{2} - \frac{5}{2} s^2 - \scs \left(5 - \frac{3 c^2}{2 \ccs}\right)
   \right] \right\}	- \frac{\alpha}{\pi}\,\hat{d},
\label{E10c}
\eea
where
\be
	\hat{d} = {d\over1 - 8\scs / 3} .
\label{E10d}
\ee
\esub
The bosonic contribution $(e^2/\scs\mws) [ \Awwp{(b)}(\mws) - \Awwp{(b)}(0)
]_{\sms}$ is given in Eq.~(A.6) of Ref.\cite{R1}.
In order to study the fermionic contributions, denoted by a superscript
$(f)$, we define
\bsub
\be
  B^{(f)} \equiv \frac{e^2}{\scs} \Re \left[\frac{\Awwp{(f)}(\mws) -
  \Awwp{(f)}(0)}{\mws}
  \right]_{\sms} ,
\label{E11a}
\ee
and write
\be
	B^{(f)} = B^{(f)}_{0} + B^{(f)}_{\rm QCD} ,
\label{E11b}
\ee
where $B^{(f)}_0$ and $B^{(f)}_{\rm QCD}$ stand for the \Ord{\alpha} and
\Ord{\alpha \alpha_s} corrections.
If we neglect very small terms proportional to $\alpha\,m^2_q/\mws$ ($q =
d,s,b$), the mixing angles in the quark sector are irrelevant~\cite{R40}
and to zeroth order in $\alpha_s$ we have (cf.\ Eq.~(A.5) of
Ref.\cite{R1})
\bea
  B_0^{(f)} &=& \frac{\alpha}{2 \pi \scs} \left\{
    2 \left(\ln{c^2} - \frac{5}{3}\right) + \frac{\ln\omega_t}{2}  +
    \frac{\omega_t}{8}(1 + 2 \omega_t) + \frac{(\omega_t -1)^2}{2}
    \left(1+{\omega_t\over2}\right) \ln\left(1-\frac{1}{\omega_t}\right)
    \right\},
 \nonumber \\*
{} & {} & {}\label{E11c}
\eea
where we have included the lepton and quark contributions and
$\omega_t = \mts/\mws$.
As a refinement, in Appendix~C we discuss the contributions of
\Ord{m^2/\mws}, where $m$ stands for a lepton or quark mass other than
\mt.
As these terms are very small, we may neglect in their evaluation the
squares of the mixing angles.
In that case one obtains a sum of isodoublet contributions, which, for
arbitrary masses, is given in Appendix~C.
One finds that the corrections of \Ord{m^2/\mws} to $B_0^{(f)}$ are
indeed very small, of \Ord{10^{-5}}.

The contribution of \Ord{\alpha \alpha_s} in Eq.~(\ref{E11b}) is given by
(see Appendix~A):
\bea
  B_{\rm QCD}^{(f)} &=& \frac{\alpha}{4 \pi \scs} \left\{
  	\frac{2 \alphah_s(\mz)}{\pi} \left( \ln{c^2} + 4 \zeta(3) -
	\frac{55}{12} \right) \right. \nonumber\\*
 & &   \left.
      + \frac{\alphah(\mt)}{\pi} \left[ \ln{c^2} + 4 \zeta(3) - \frac{55}{12}
      - 4 \omega_t \left( F_1\left({1\over\omega_t}\right) - F_1(0) \right) +
      \ln\omega_t\right]\right\} ,
\label{E11d}
\eea
\esub
where $F_1(x)$ is defined in Ref.\cite{R25}. In Eq.~(\ref{E11d}) we have
neglected all quark masses other than \mt.
The first term in Eq.~(\ref{E11d}) arises from the $(u,d)$ and $(c,s)$
isodoublets, while the second is associated with the $(t,b)$ doublet.
In Eqs.~(\ref{E10c}--d,\ref{E11d}) the mass scale of $\alphah_s$ has been
chosen according to the prescription explained at the end of Section~2.

The asymptotic behavior of \drcarw\ for large \mt\ can be obtained from
Eqs.~(\ref{E9c},\ref{E10c}--d,\ref{E11c}--d) and is given by
\be
  \drcarw \sim \frac{\alpha}{6 \pi \scs} \left(1 + \frac{\alphah_s(\mt)}{\pi}
  \right)
  \ln\frac{\mt}{\mz} \qquad\qquad (\mt \gg \mz) .
\label{E12}
\ee
This exhibits a smaller coefficient than the corresponding expression in
Ref.\cite{R12}, a feature that is due to a partial cancellation with the
finite counterterm $-(\alpha/\pi) d$ in Eq.~(\ref{E10a}).
The asymptotic behavior for $\mh \gg \mz$ is the same as in
Ref.\cite{R12}, namely
\be
   \drcarw \sim \frac{\alpha}{24 \pi \scs}
   \ln\frac{\mh}{\mz} \qquad\qquad (\mh \gg \mz) .
\label{E13}
\ee

We now turn our attention to \drcar. When the top decoupling is
implemented according to Section~2, the expression for \drcar\ becomes
(see Appendix~B)
\bea
   \drcar &=& \drcarw - \frac{\ehs \left[1 - \drcarw -
   (\alpha/\pi)\hat{d}\,\right]}{\scs\,\mzs}\Re\left[\frac{\Aww(\mws)}{\ccs} -
   \Azz(\mzs) \right]_{\sms}
   + \frac{\alpha}{\pi}\,\frac{\scs}{\ccs}\,d ,
\label{E14}
\eea
where $d$ and $\hat{d}$ are defined in Eqs.~(\ref{E9c},\ref{E10d}) and
$\Azz(q^2)$ is the unrenormalized $ZZ$ transverse self energy with the
coupling $e^2/\scs$ factored out~\cite{R1}.
Except for the terms involving $d$ and $\hat{d}$, which arise from the
finite counterterms in Eqs.~(\ref{E7d},\ref{E9a}) associated with the
decoupling of the top quark, Eq.~(\ref{E14}) has the same form as Eq.~(15b)
of Ref.\cite{R1}.
It is understood, however, that \drcarw, \ehs, and \scs\ in Eq.~(\ref{E14})
are defined according to the prescriptions of the present paper,
namely Eqs.~(\ref{E10c},\ref{E8c},\ref{E2}).
In Appendix~B we show how Eqs.~(\ref{E10a}--c,\ref{E14}) can be derived
from the results of Ref.\cite{R1} by neglecting very small contributions
of \Ord{\alpha^2} without logarithmic or $\mts/\mzs$ enhancements, as
well as terms of \Ord{\alpha^3}.
As pointed out in Ref.\cite{R1}, if one neglects also subleading
corrections of \Ord{(\alpha/\pi \scs) x_t}, with $x_t$ being defined in
Eq.~(\ref{E17b}), one can replace
$\ehs \left[1-\drcarw -(\alpha/\pi)\hat{d}\,\right] \rightarrow e^2$
in the second term of Eq.~(\ref{E14}).
As in Ref.\cite{R1}, we have retained such subleading terms
in Eq.~(\ref{E14}) because the resulting expression describes very
accurately the resummation of one-loop effects, which is particularly
simple in the \ms\ framework.

We now turn our attention to the evaluation of the self energies in the
second term of Eq.~(\ref{E14}).
The bosonic contribution
$ (e^2/\scs \mzs)\Re\left[{\Awwp{(b)}(\mws)}/{\ccs}
- \Azzp{(b)}(\mzs) \right]_{\sms} $
is given in Eq.~(A.9) of Ref.\cite{R1}.
To study the fermionic contributions, we write
\bsub
\be
 C^{(f)} \equiv \frac{e^2}{\scs\mzs}\Re\left[\frac{\Awwp{(f)}(\mws)}{\ccs} -
    \Azzp{(f)}(\mzs) \right]_{\sms} ,
\label{E15a}
\ee
and expand
\be
  C^{(f)} = C_{0}^{(f)} + C_{\rm QCD}^{(f)} ,
\label{E15b}
\ee
where $C_{0}^{(f)}$ and $C_{\rm QCD}^{(f)}$ are the \Ord{\alpha} and
\Ord{\alpha \alpha_s} corrections, respectively.
If the small terms proportional to $\alpha\,m_q^2/\mws$ ($q=d, s, b$)
are neglected, the mixing angles are once more irrelevant and, to
zeroth order in $\alpha_s$, we have (cf.\  Eq.~(A.8) of Ref.\cite{R1})
\bea
  C_0^{(f)} &=&
  \frac{\alpha}{2\pi\scs}\, \frac{c^2}{\ccs}
  \left\{
  2\left(\ln{c^2} - \frac{5}{3}\right) + \frac{\ln\omega_t}{2}
  + \frac{\omega_t}{8}(1 + 2 \omega_t)
  + \frac{(\omega_t -1)^2}{2} (1 + \frac{\omega_t}{2})
  \ln\left(1-\frac{1}{\omega_t}\right)
  \right\} \nonumber\\*
  & & + \frac{\alpha}{2\pi\scs\ccs}
  \left\{
  \frac{5}{3}\left(\frac{7}{4}-\frac{10}{3}\scs + \frac{40}{9}\hat{s}^4\right)
  - \frac{1}{8}\left[1 + \left(1-\frac{8}{3}\,\scs\right)^2\right]
  \left[\ln\mu_t + \frac{1}{3} +
  2(1+2\mu_t)
  \right.\right.\nonumber\\*
  & & \left.\vphantom{\left[1 + \left(1-\frac{8}{3}\,\scs\right)^2\right]}
 \left.\vphantom{1\over3}
 \times (\Lambda(D_t)-1)\right]\right\}
  + \frac{3 \alpha}{4\pi \scs\ccs} \mu_t \left(\frac{1}{4} + \Lambda(D_t) -1
 \right) ,
\label{E15c}
\eea
where $\mu_t = \mts/\mzs$, $D_t = 4 \mu_t - 1$, and
$\Lambda(D) = D^{1/2} \tan^{-1}(D^{-1/2})$ for $D > 0$.
In analogy with our discussion of $B_0^{(f)}$, in Appendix~C we give the
expression for $C_0^{(f)}$ for arbitrary fermion masses in the
approximation of neglecting the squares of mixing angles.
We find that the corrections of \Ord{m^2/\mws} to $C_0^{(f)}$ are
also of \Ord{10^{-5}}.

The contribution of \Ord{\alpha \alpha_s} in Eq.~(\ref{E15b}) is given by
\bea
 C^{(f)}_{\rm QCD} &=&
   \frac{\alpha}{4\pi\scs\ccs} \left\{
   \frac{2 \alphah_s(\mz)}{\pi} \left[
    c^2 \ln{c^2} + \left(4 \zeta(3) - \frac{55}{12}\right)
   \left(-s^2 + 2 \scs \left(1 -
    \frac{10}{9} \scs\right) \right)
   \right]
   \right. \nonumber\\*
   & &
   + \frac{\alphah_s(\mt)}{\pi} \left[
   \vphantom{\left(1-\frac{8}{3}\scs\right)^2}
    c^2 \ln{c^2} +
    \left(4 \zeta(3) - \frac{55}{12}\right)
   \left(-s^2 + 2 \scs \left(1-\frac{10}{9} \scs\right)\right)
    - c^2 \left(4 \omega_t F_1\left({1\over\omega_t}\right)
   \right.\right.
   \nonumber\\*
   & & \left.\left.\left.\vphantom{\left({1\over\omega_t}\right)}
   - \ln\omega_t\right)
   +  \left(1-\frac{8}{3}\scs\right)^2 \left(\mu_t
   V_1\left({1\over4\mu_t}\right) - {\ln\mu_t\over4}\right)
   +\mu_t A_1\left({1\over4\mu_t}\right) - {\ln\mu_t\over4}
   \right]\right\},
   \nonumber\\*
  & & \label{E15d}
\eea
\esub
where the functions $F_1(x)$, $V_1(r)$, and $A_1(r)$ are defined in
Ref.\cite{R25}.

The leading asymptotic behavior of $C_0^{(f)}$ for large \mt\ is
$(3\alpha/16\pi\scs\ccs)(\mts/\mzs)$, which arises from the last term of
Eq.~(\ref{E15c}).
In the \Ord{\alpha \alpha_s} corrections the leading contribution is
contained in the combination
$ \left(\alpha \alphah_s(\mt)/4 \pi^2 \scs\ccs\right) \left[
 \mu_t A_1(1/(4\mu_t)) - 4 c^2 \omega_t F_1(1/\omega_t)\right] $,
which asymptotically becomes $-\left(\alpha \alphah_s(\mt)/8\pi^2
\scs\ccs\right)(\pi^2/3+1)\left(\mts/\mzs\right)$.
Combining these contributions, inserting the result in Eq.~(\ref{E14})
and neglecting there the subleading \Ord{\alpha^2} contributions, we
have
\bsub
\be
 \drcar \sim - \frac{3 \alpha}{16 \pi \scs \ccs}\, \frac{\mts}{\mzs}
  \left[ 1 - \frac{2\alphah_s(\mt)}{3\pi} \left(\frac{\pi^2}{3} +
  1\right) \right]\qquad\qquad (\mt \gg \mz) ,
\label{E16a}
\ee
while the leading asymptotic behavior for large \mh\ is the same as in
Ref.\cite{R12}, namely
\be
 \drcar \sim \frac{\alpha}{2\pi\scs\ccs}\left(\frac{5}{6} - \frac{3}{4}\ccs
 \right) \ln\frac{\mh}{\mz}\qquad\qquad (\mh \gg \mz) .
\label{E16b}
\ee
The term involving $\alphah_s$ in the last factor of Eq.~(\ref{E16a})
represents the most important \Ord{\alpha \alpha_s} correction.
Indeed, the contribution of Eq.~(\ref{E16a}) depends very sensitively on
\mt\ and the coefficient of $\alphah_s/\pi$ in the last factor, namely
$-2.86$, is quite large.
The presence of this $\alphah_s$ correction induces an increase in the
value of \mt\ read from experiments of approximately
\be
  \frac{\Delta\mt}{\mt} \approx
  \frac{\alphah_s(\mt)}{3\pi}\left(\frac{\pi^2}{3}+1\right) \approx
  0.455\, \alphah_s(\mt) ,
\label{E16c}
\ee
\esub
which amounts to $\Delta\mt \approx (4.9$, 7.5, 9.6, 11.7)~GeV for
$\mt = \mz$, 150, 200, 250~GeV.
It should be stressed that, as is obvious from the structure of
Eq.~(\ref{E16a}), these results depend sensitively on the precise
definition of \mt.
The quantity that appears in Eq.~(\ref{E16a}) and the various expressions
of this paper is the zero of the real part of the inverse propagator.
In the literature it is variously referred to as the ``physical,''
``on-shell,'' or ``dressed'' mass.
In the approximation of neglecting the $s$ dependence of the imaginary
part of the top-quark self energy, it coincides with the real part of
the complex pole position~\cite{R13,R14,R15}.
It is also the mass that occurs in the Balmer formula for the toponium
levels in the non-relativistic bound-state picture and the parameter
that governs the start of the $t \bar{t}$ cut in perturbation
theory~\cite{R43}.
All the recent calculations of \Ord{\alpha \alpha_s} and $t \bar{t}$
production~\cite{R22,R23,R24,R25,R26,R27,R28,R29,R30,R31}
employ this definition or slight modifications thereof.
It is worthwhile to notice that the $\Ord{\alphah_s}$ corrections become much
smaller if one employs other definitions of mass~\cite{R44}.
For example, \mt\ and the running mass, $\hat{m}_t(\hat{m}_t)$, are
related by $\mt = \hat{m}_t(\hat{m}_t)\left[ 1 + 4 \alphah_s/(3 \pi) +
\Ord{\alphah_s^2} \right]$~\cite{R43}.
Inserting this into Eq.~(\ref{E16a}), we get a contribution involving
$(\hat{m}_t(\hat{m}_t))^2 \left[1 - 2(\alphah_s/\pi)(\pi^2/9-1)\right]$
and we see that the coefficient of $\alpha_s/\pi$ has changed from $-2.86$
to $-0.19$.
Similarly, if one expresses the corrections in terms of the
Georgi--Politzer mass, $M(-m^2)$~\cite{R45}, which is gauge-dependent and
usually evaluated in the Landau gauge, the coefficient of
$\alphah_s/\pi$ becomes even smaller, namely +0.09.
Because the perturbative evaluation of the $(t,b)$-isodoublet loops
involves high mass scales, of \Ord{\mt} or \Ord{\mz}, both the on-shell and
$\hat{m}_t(\hat{m}_t)$ definitions are in principle suitable, although
the former is the natural choice in the DR
approach~\cite{R23,R24,R25,R30,R31}.
The relevant question, of course, is what mass parametrization is more
adequate to describe the physical issues at hand, namely the production
and detection of the top quark.
In this connection it also appears that the on-shell mass is the most
appropriate parameter because, in the propagation of $t$ and $\bar{t}$
between the ``production'' and ``decay'' vertices, configurations near
the ``mass-shell'' will be greatly enhanced kinematically (resonance
effect).
Another consideration, of a more practical nature, is that the \mt\
parametrization of the radiative corrections should be consistent with
the one employed in the calculation of $t\bar{t}$
production~\cite{R27,R28,R29} and, as mentioned before,
this is again the ``pole'' or ``on-shell'' definition.

Returning to the evaluation of \drcar, we must still consider the
irreducible contributions of \Ord{\alpha^2(\mts/\mws)^2}.
As mentioned before, these can be gleaned from the two-loop irreducible
corrections to the $\rho$ parameter.
The very recent work of Ref.\cite{R7} leads to a significant change in
the magnitude of these corrections.
Indeed, these authors find for the leading high-\mt\ contributions to the
$\rho$ parameter an expansion of the form
\bsub
\be
 \frac{1}{\rho} = 1 - x_t \left( 1 + R\left({\mh\over\mt}\right)
 \frac{x_t}{3}\right) \equiv
 1 - \Drhob ,
\label{E17a}
\ee
where
\be
	x_t = \frac{3 \Gmu \mts}{8 \pi^2\sqrt{2}}
\label{E17b}
\ee
\esub
is the one-loop term~\cite{R46}
and $R$ is a negative function of $\mh/\mt$.
When $\mh/\mt =0$, $R$ equals $19-2 \pi^2 \approx -0.7392$, the result of
Ref.\cite{R6}, but as $\mh/\mt$ increases, $R$ rapidly becomes more
negative, reaching a minimum of $\approx -11.8$ for $\mh/\mt \approx
5.8$.
In current discussions $\mh/\mt$ ranges from $\approx 0.24$
(corresponding to $\mh \approx 60 \GeV$ and $\mt \approx 250 \GeV$) to
$\approx 11$ (corresponding to $\mh \approx 1 \TeV$ and $\mt \approx
91 \GeV$).
Noting that $R(0.24) \approx -3$ and $R(11) \approx -10$, it is clear
that, although Refs.\cite{R6} and \cite{R7} agree in the limit
$\mh/\mt =0$, for realistic values of this ratio the results of
Ref.\cite{R7} tell us that these corrections are considerably larger in
magnitude.

Calling $\drcar^{(1)}$  the ``one-loop'' expression for \drcar\ given in
Eq.~(\ref{E14}), we include the two-loop irreducible contributions by
writing
\bsub
\be
	\drcar = \drcar^{(1)} - {1\over3}R\left({\mh\over\mt}\right)x_t^2\,
	\frac{\left(1-\drcar^{(1)}\right)^2}{1-\drcarw} .
\label{E18a}
\ee
The rationale is the following.
In the \ms\ scheme the $\rho$ parameter is naturally identified with
$\hat{\rho} \equiv c^2/\ccs$ and, from Eqs.~(\ref{E2},\ref{E3}), we see that
(cf.\ Eqs.~(17a,b) of Ref.\cite{R1})
\be
    \hat{\rho} = {1 - \drcar\over1-\drcarw} .
\label{E18b}
\ee
\esub
Neglecting very small terms of order $(R\,x_t^2/3)^2$, one indeed verifies
that when the second term in Eq.~(\ref{E18a}) is inserted in Eq.~(\ref{E18b}),
it leads to an additional contribution of $-R(\mh/\mt)x_t^2/3$ to
$1/\hat{\rho}$, in conformity with Eq.~(\ref{E17a}).

We have given all the elements necessary to evaluate the basic radiative
corrections \drcarw\ (cf.\ Eq.~(\ref{E10c})) and \drcar\
(cf.\ Eqs.~(\ref{E14},\ref{E18a})) including \Ord{\alpha \alpha_s} corrections.
In conjunction with
\bsub
\be
       \scs = \frac{1}{2} \left\{
                1 - \left[ 1 - \frac{4 A^2}{\mzs(1-\drcar)} \right]^{1/2}
		           \right\} ,
\label{E19a}
\ee
which follows from Eq.~(\ref{E3}), \drcar\ can be employed to calculate
$\scs \equiv \sincarmz$ in terms of the accurately known quantities \Gmu,
\mz, and $\alpha$, as a function of \mt\ and \mh.
The parameter $s^2 \equiv 1 - \mws/\mzs$ can be computed from (cf.\ Eq.~(19) of
Ref.\cite{R1})
\be
       s^2 = \scs \left( 1 - \frac{\ccs}{\scs}\,
                  \frac{\drcarw - \drcar}{1 - \drcarw} \right)
\label{E19b}
\ee
and \dr\ (cf.\ Eq.~(22) of Ref.\cite{R1}) from
\be
       \dr = \drcarw - \frac{\ccs}{\scs}\, \frac{\drcarw-\drcar}
               { 1-\left(\ccs/\scs\right)\left(\drcarw-\drcar\right)/
                \left(1-\drcarw\right)} .
\label{E19c}
\ee
Alternatively, writing $\hat{\rho} \equiv 1/(1-\Delta\rhoh)$ we have (cf.\
Eqs.~(17a,b,20) of Ref.\cite{R1})
\bea
   \Delta\rhoh &=& \frac{\drcarw - \drcar}{1 - \drcar}\, ,
   \label{E19d} \\*
   1 - \dr &=& \left( 1 + \frac{c^2}{s^2} \Delta\rhoh\right)
   \left(1 - \drcarw\right) .
   \label{E19e}
\eea
Again, the \mw--\mz\ interdependence can be expressed in two equivalent
forms (cf.\ Eqs.~(24,25) of Ref.\cite{R1}),
\bea
        \frac{\mws}{\mzs} &=& \frac{1}{2} \left\{
         1 + \left[ 1 - \frac{4 A^2}{\mzs(1-\dr)} \right]^{1/2}
                            \right\} ,
        \label{E19f}\\*
        \frac{\mws}{\mzs} &=& \frac{\rhoh}{2} \left\{
         1 + \left[ 1 - \frac{4 A^2}{\mzs\rhoh(1-\drcarw)} \right]^{1/2}
                            \right\} .
        \label{E19g}
\eea
\esub
The $W$ mass can be evaluated from (\ref{E19b}), or (\ref{E19f}), or
(\ref{E19g}).
Equations~(\ref{E19a}--g) have the same structure as in Ref.\cite{R1} because
they follow from the same basic relations, namely Eqs.~(\ref{E1}--\ref{E3}).
In this paper, however, the explicit evaluation of \drcarw\ and \drcar\
via Eqs.~(\ref{E10c},\ref{E14},  \ref{E18a}) is somewhat different because
we have included the \Ord{\alpha \alpha_s} corrections, updated the
contributions of $e^2\Re\left(\piggp{5}(0) - \piggp{5}(\mzs)\right)$
and the two-loop irreducible parts, and implemented the decoupling of the
top quark.
By inference the same holds true for \dr\ and $\Delta\rhoh$ when they are
evaluated from \drcarw\ and \drcar\ via Eqs.~(\ref{E19c}) and
(\ref{E19d}), respectively.
It should be observed moreover that, except for very small effects of
\Ord{\alpha^2}, i.e.\  of the same order as those we have neglected, the
decoupling of the top quark should not affect physical
observables such as the radiative correction \dr\ and the predicted value
of \mw.

As shown in Appendix~C, the corrections of \Ord{m_f^2/\mws} to \dr\ are
also very small.
However, they are enhanced relative to those of \drcarw\ and \drcar, and
for $\mz \leq \mt \leq 250 \GeV$ they vary from $\approx -7\times10^{-5}$
to $\approx 8\times10^{-5}$.
\section{Residual Parts of \drcarw, \drcar, and \dr}
It is a simple matter to derive expressions relating \mw\ and \mz\ to
\Gmu, \scs, \ccs, and \ehs\ (rather than $e^2$).
To see this, we write Eq.~(\ref{E2}) in the form
\bsub
\be
  \scs = \frac{\pi \alpha}{\sqrt{2} \Gmu \mws}\,
  \frac{1}{1 - (\alpha/\pi)\Delta_\gamma
    -\left(\drcarw-(\alpha/\pi)\Delta_\gamma\right)}.
\label{E20a}
\ee
Factoring out $\left(1 - (\alpha/\pi)\Delta_{\gamma}\right)$ and recalling
Eq.~(\ref{E8c}), we have
\be
  \scs = \frac{\pi \alphah}{\sqrt{2} \Gmu \mws}\,
    \frac{1}{1 - (\drcarw)_{\rm res}},
\label{E20b}
\ee
where
\be
 (\drcarw)_{\rm res}  =
\left(\drcarw - \frac{\alpha}{\pi}\Delta_{\gamma} \right)
 \frac{\ehs}{e^2} .
\label{E20c}
\ee
\esub
The correction $(\drcarw)_{\rm res}$ represents the ``residual part'' of
\drcarw\ after we have subtracted the large contribution $(\alpha/\pi)
\Delta_{\gamma}$, evaluated with the coupling $\ehs$ rather than
$e^2$.
As $(\drcarw)_{\rm res} \ll \drcarw$, we see that the dominant part of
\drcarw\ can be absorbed by employing \alphah\ rather than $\alpha$ as
zeroth-order coupling.

Starting with Eq.~(\ref{E3}), the analogous argument leads to
\bsub
\be
 \scs \ccs = \frac{\pi \alphah}{\sqrt{2} \Gmu \mzs}\,
   \frac{1}{1 - (\drcar)_{\rm res}} ,
\label{E21a}
\ee
where
\be
   (\drcar)_{\rm res} = \left(\drcar - \frac{\alpha}{\pi}
  \Delta_{\gamma}\right)
   \frac{\ehs}{e^2}.
\label{E21b}
\ee
\esub
Again $(\drcar)_{\rm res}$ is the ``residual part'' of \drcar.
In connection with the inclusion of the two-loop irreducible
contributions of \Ord{\alpha^2(\mts/\mws)^2} (cf.\ Eq.~(\ref{E18a})),
it is easy to see that the ``one-loop'' $\left(\drcar^{(1)}\right)_{\rm res}$
can be obtained by replacing $\drcarw \rightarrow (\drcarw)_{\rm res}$,
$(\alpha/\pi)\hat{d} \rightarrow (\alphah/\pi) \hat{d}$, and
$(\alpha/\pi) d \rightarrow (\alphah/\pi) d$ on the r.h.s.\ of
Eq.~(\ref{E14}), and the final $(\drcar)_{\rm res}$ follows by substituting
$\drcar^{(1)} \rightarrow \left(\drcar^{(1)}\right)_{\rm res}$ and
$\drcarw \rightarrow (\drcarw)_{\rm res}$ on the r.h.s.\ of Eq.~(\ref{E18a}).
Of course, once \drcar\ is known, one can directly use Eq.~(\ref{E21b})
for the numerical evaluation of $(\drcar)_{\rm res}$.

The corresponding expression from Eq.~(\ref{E1}) is
\bsub
\be
 s^2 = \frac{\pi \alphah}{\sqrt{2} \Gmu \mws}\,
    \frac{1}{1 - (\dr)_{\rm res}} ,
\label{E22a}
\ee
where
\be
   (\dr)_{\rm res} = \left(\dr - \frac{\alpha}{\pi} \Delta_{\gamma}\right)
      \frac{\ehs}{e^2}
\label{E22b}
\ee
\esub
is the ``residual part'' of \dr.
One readily finds that this quantity can be obtained by simply
substituting $\drcarw \rightarrow (\drcarw)_{\rm res}$ and
$\drcar \rightarrow (\drcar)_{\rm res}$ everywhere on the r.h.s.\ of
Eq.~(\ref{E19c}).
Of course, Eq.~(\ref{E22b}) can be directly used for numerical
evaluations.

As illustrations, for $\mh = 250 \GeV$ and $\mt = 150 \GeV$ we have:
$\drcarw = 7.02\times10^{-2}$, $\drcar = 6.34\times10^{-2}$, and $\dr =
4.74\times10^{-2}$, while $(\drcarw)_{\rm res} = 3.6\times10^{-3}$,
$(\drcar)_{\rm res} = -3.6\times10^{-3}$, and
$(\dr)_{\rm res} = -2.08\times10^{-2}$.
The corresponding values for $\mh= 250 \GeV$ and $\mt = 200 \GeV$ are
$\drcarw = 7.08\times10^{-2}$, $\drcar = 5.88\times10^{-2}$, and $\dr =
2.90\times10^{-2}$, while
$(\drcarw)_{\rm res} = 4.3\times10^{-3}$, $(\drcar)_{\rm res}
 = -8.5\times10^{-3}$, and
$(\dr)_{\rm res} = -4.05\times10^{-2}$.
Unlike $(\drcarw)_{\rm res}$ or $(\drcar)_{\rm res}$, $(\dr)_{\rm res}$ is
quite
large for $\mt \approx 200\GeV$.

A correction similar to $(\dr)_{\rm res}$ has been recently employed in
Ref.\cite{R20}.
The two quantities are, however, not identical because \alphah, defined
in the \ms\ scheme, differs somewhat from the effective parameter
$\alpha(\mz) = (128.8)^{-1}$ used in that work.
This illustrates the rather obvious but important fact that running
couplings are scheme-dependent.
\section{\dr\ in the On-Shell Scheme}
In this section we discuss the incorporation of the perturbative \Ord{\alpha
\alpha_s} corrections to \dr~\cite{R2,R3} in the on-shell scheme of
renormalization~\cite{R2}.
We follow the formulation proposed recently in Ref.\cite{R8}, based on
the expression
\bsub
\be
  \dr = \da - \frac{c^2}{s^2} \Drhob\, (1 - \da) + (\dr)_{\rm rem} ,
\label{E23a}
\ee
where $\da = 0.0597 \pm 0.0009$ represents the contribution of the
charged leptons and the first five quark flavors to the photon
vacuum-polarization function evaluated at $q^2
= \mzs$, i.e.\  $e^2\Re\left(\pigg(0) - \pigg(\mzs)\right)$,
\be
   \Drhob = x_t \left( 1 + R\left({\mh\over\mt}\right) {x_t\over3}\right),
\label{E23b}
\ee
\esub
$x_t$ is defined in Eq.~(\ref{E17b}), and $R(\mh/\mt)$~\cite{R7} is the
function discussed after Eq.~(\ref{E17a}).
The result for \da\ quoted above includes the recent calculation of
the first five quark flavor contribution~\cite{R10} and, for this reason,
it slightly differs from the central value of $0.0602$ employed in
Ref.\cite{R8}.
The second term in Eq.~(\ref{E23a}) involves the leading \mt-dependent
correction \Drhob\ to $1 - 1/\rho$ (cf.\ Eq.~(\ref{E17a})) and we see
that, in the case of \dr, it is enhanced by a factor $c^2/s^2$.
Its importance for large \mt\ in the \mw--\mz\ interdependence was
pointed out in 1980, in the work of
W.~J.~Marciano and one of us (A.S.)~\cite{R40}.
Since that time, this potential effect has been discussed by several
authors~\cite{R4,R5,R22,R23,R24,R25,R26,R47,R48}.

The separation into leading contributions (the first two terms in
Eq.~(\ref{E23a})) and a ``remainder'' $(\dr)_{\rm rem}$ is the same as
was proposed in Refs.\cite{R4,R5,R26},
except that we have included the recent results of
Ref.\cite{R7} on the two-loop contribution to \Drhob.
It is important to note that $(\dr)_{\rm rem}$ differs from the quantity
$(\dr)_{\rm res}$ introduced at the end of Section~3.
Whereas in the latter we subtract the large logarithmic corrections, in
$(\dr)_{\rm rem}$ we also exclude the leading \mt-dependent contributions.
The formulation of Ref.\cite{R8} provides also a very specific
prescription to calculate $(\dr)_{\rm rem}$, namely
\bsub
\be
   (\dr)_{\rm rem} =
      \dr^{(1)} - \da + \frac{c^2}{s^2} X + \frac{c^2}{s^2} (\tilde{x}_t -
      X) \frac{\sqrt{2} \Gmu \mws (1 - \da) s^2}{\pi \alpha} ,
\label{E24a}
\ee
where $\dr^{(1)}$ is the familiar one-loop expression of Ref.\cite{R2},
expressed in terms of $\alpha$ and $\alpha/s^2$ as expansion parameters,
\bea
	\tilde{x}_t &=& \frac{3 \alpha}{16 \pi s^2}\, \frac{\mts}{\mws} ,
	\label{E24b} \\*
	X &=& \frac{e^2}{s^2} Re\left[
	 \frac{\Aww(\mws)}{\mws} - \frac{\Azz(\mzs)}{\mzs}\right]_{\rm fin} ,
	\label{E24c}
\eea
\esub
and the subscript $\rm fin$ means ``finite part'', i.e.\  that
the pole terms have been subtracted and $\mu'$ has been set equal to \mz.
We also note that $X$ is a gauge-invariant quantity.
In Eq.~(\ref{E24c}) we follow the notation of Ref.\cite{R1}, which differs
from that of Refs.\cite{R2,R40} in that an explicit coupling $e^2/s^2$
has been factored out in the \Aww\ and \Azz\ self energies.
The term $-\da$ in Eq.~(\ref{E24a}) subtracts from $\dr^{(1)}$ the large
logarithmic corrections which are included as part of the leading
contributions in Eq.~(\ref{E23a}).
The quantity $(c^2/s^2) X$ subtracts another important part of
$\dr^{(1)}$, which is then treated more accurately in the second term of
Eq.~(\ref{E23a}) and the third term of Eq.~(\ref{E24a}), according to the
following prescription.
Decomposing $-(c^2/s^2) X = -(c^2/s^2) \tilde{x}_t + (c^2/s^2)
(\tilde{x}_t - X)$, the dominant part, $-(c^2/s^2) \tilde{x}_t$, is
included in the second term of Eq.~(\ref{E23a}) with the effective
coupling constant changed according to
$\alpha/s^2 \rightarrow \sqrt{2} \Gmu \mws (1-\da)/\pi$.
The non-dominant part, $(c^2/s^2)(\tilde{x}_t - X)$, is treated with the
same coupling modification but it is retained as part of $(\dr)_{\rm rem}$
(third term of Eq.~(\ref{E24a})).
The rationale for this treatment of
$(c^2/s^2)(\tilde{x}_t - X)$ was explained in Refs.\cite{R8,R49} and
reflects the fact that a careful analysis of the resummation of one-loop
effects leads to an expression of the form of Eq.~(\ref{E23a}) in which
$X$, rather than its dominant part $x_t$, is multiplied by
$\sqrt{2} \Gmu s^2 \mws (1-\da)/(\pi \alpha) = (1 - \da)/(1-\dr)$.
This is natural because, since \dr\ is the radiative correction in the
relation between \mw, \mz, \Gmu, and $\alpha$, it should
involve quantities evaluated at $q^2 = \mzs$ and $q^2 = \mws$ such as
$X$, rather than amplitudes evaluated at $q^2 = 0$ such as \Drhob.
We also point out that the neglect of this effect in Eq.~(\ref{E24a}),
namely the replacement $\sqrt{2} \Gmu \mws s^2 (1-\da)/(\pi \alpha)
\rightarrow1$ in the last term, would induce a change
$\approx (c^2/s^2) (\tilde{x}_t - X) (c^2/s^2) \Drhob =
\Ord{(\alpha/\pi s^2)(c^2/s^2)^2 \Drhob}$;
 although formally subleading, this is enhanced by two powers of
$c^2/s^2$ and is, therefore, significantly larger than  the expected
theoretical error.
It was already pointed out in Ref.\cite{R8} that
Eqs.~(\ref{E23a},\ref{E24a}) include correctly not only the leading terms
of \Ord{\alpha^2(\mts/\mws)^2} and \Ord{\alpha^2 \ln^2(\mz/\mf)}, where
\mf\ is a generic fermion mass, but also the subleading contributions of
\Ord{\alpha^2 \ln(\mz/\mf)}.

We now turn our attention to the incorporation of the perturbative
\Ord{\alpha\alpha_s} contributions.
We first consider Eq.~(\ref{E24a}).
The amplitudes modified by the QCD corrections are $X^{(f)}$ in the last
term and the self-energy contributions
$(e^2/s^2\mws) Re\left[\Awwp{(f)}(\mws)\right.$
$\left.-\Awwp{(f)}(0)\right]_{\rm fin}$ and
$e^2\left[\piggp{f}(0)\right]_{\rm fin}$ contained in $\dr^{(1)}$,
where the superscript $(f)$ denotes again fermionic contributions.
The first two are obtained from
$C^{(f)}$ (cf.\ Eqs.~({\ref{E15a}--d})) and
$B^{(f)}$ (cf.\ Eqs.~({\ref{E11a}--d})), respectively, by simply
changing everywhere $\scs\rightarrow s^2$ and $\ccs\rightarrow c^2$.
In particular, their QCD corrections are derived from Eqs.~(\ref{E15d})
and (\ref{E11d}), respectively.
The correction $e^2\left[\piggp{f}(0)\right]_{\rm fin}$ can be read from
Eq.~(\ref{E7a}) by removing the pole terms, setting $\mu'=\mz$, and
substituting $\alphah\rightarrow\alpha$.
It is not necessary to consider $(c^2/s^2)X^{(f)}$ in the third term
of Eq.~(\ref{E24a}) because, as explained before, it cancels an identical
contribution in $\Delta r^{(1)}$.
The value of $\Delta\alpha$ is not modified, as QCD corrections have
already been taken into account in its evaluation.

In the above discussion, the quantities $\tilde x_t$ in Eq.~(\ref{E24a})
and $\Delta\bar\rho$ in Eq.~(\ref{E23a}) have not been altered, so that,
except for $\Delta\alpha$, all the QCD corrections are contained in
$(\Delta r)_{\rm rem}$.
We may wish, however, to incorporate the leading QCD corrections in the
second term of Eq.~(\ref{E23a}).
To achieve this, we subtract them from Eq.~(\ref{E24a}) by replacing
\bsub
\be
\tilde x_t\rightarrow\tilde x_t'={3\alpha\over16\pi s^2}\,
{\mts\over\mws}\left[1-{2\alphah_s(\mt)\over3\pi}
\left({\pi^2\over3}+1\right)\right]
\label{E25a}
\ee
in the last term of that equation and, at the same time, we substitute
\be
\Delta\bar\rho\rightarrow\Delta\bar\rho'=x_t
\left[1-{2\alphah_s(\mt)\over3\pi}\left({\pi^2\over3}+1\right)
+R\left({\mh\over\mt}\right){x_t\over3}\right]
\label{E25b}
\ee
\esub
in Eqs.~(\ref{E23a},b).
The overall evaluation of $\Delta r$ is, of course, the same whether we
employ $\Delta\bar\rho$ in Eq.~(\ref{E23a}) and $\tilde x_t$ in
Eq.~(\ref{E24a}) or the modified quantities, $\Delta\bar\rho'$ and
$\tilde x_t'$.
In the second formulation, however, the first two terms of Eq.~(\ref{E23a}),
with $\Delta\bar\rho\rightarrow\Delta\bar\rho'$, describe more accurately
the leading \mt-dependent corrections.

The contributions of \Ord{\alpha^2\alphah_s, \alpha\alphah_s^2} are unknown
at the present time and for this reason we have not made any attempt to
include them.
However, the structure of Eq.~(\ref{E25b}) gives a hint about what their
magnitude might be.
Suppose, for example, that the leading QCD effects are always very small
when the electroweak corrections are expressed in terms of the running
mass, $\hat m_t\left(\hat m_t\right)$, as it happens with the
\Ord{\alpha\alphah_s} corrections.
In that hypothetical case, the discussion after Eq.~(\ref{E16c}) indicates
that the modified $\Delta\bar\rho$ parameter would be obtained
approximately by appending a factor
$\left[1-\left(2\alphah_s(\mt)/3\pi\right)(\pi^2/3+1)\right]$ to each
$x_t$ in Eq.~(\ref{E23b}).
For $\mh=600\GeV$ and $\mt=200\GeV$, the difference with Eq.~(\ref{E25b})
would lead to an aditional contribution to $\Delta r$ of $-3.4\times10^{-4}$.
This is of the same order of magnitude as the subleading terms of
\Ord{(\alpha/\pi s^2)(c^2/s^2)x_t}, discussed in Section~1.
The surprisingly large size of these possible corrections of
\Ord{(c^2/s^2)\alphah_s x_t^2} is due to the $m_t^4$ dependence and
the considerable magnitude of the function $R$.
This observation illustrates the fact that, for large \mt\ values,
the theoretical error due to unknown higher-order corrections may arise
from very different sources.
\section{Threshold Effects}

The fermionic contribution to the vacuum polarization of the
intermediate vector bosons can be expressed in terms of  the amplitudes
\bsub
\be
  \Pi_{\mu\nu}^{V,A}(q,m_1,m_2) =
   -i \int\,d^4x\,e^{i q\cdot{x}} \bra{0}T^{*}[J_{\mu}^{V,A}(x)
   J_{\nu}^{V,A\,\dagger}(0)]\ket{0},
\label{E26a}
\ee
where $T^{*}$ is the covariant time-ordered product and $J_{\mu}^{V} =
\bar{\psi}_1 \gamma_{\mu} \psi_2$ and $J_{\mu}^{A} = \bar{\psi}_1
\gamma_{\mu}\gamma_5 \psi_2$ are the vector and axial-vector currents,
constructed with the spinors fields $\psi_1$ and  $\psi_2$ endowed with
masses $m_1$ and $m_2$, respectively.
Thus, except for vector currents with $m_1 = m_2$, the conservation of
the currents is explicitly broken by mass terms.
In turn, the tensors $\Pi_{\mu\nu}^{V,A}$ have the well-known structure
\be
  \Pi_{\mu\nu}^{V,A}(q,m_1,m_2) = \Pi^{V,A}(s,m_1,m_2)\, g_{\mu\nu}
    + \lambda^{V,A}(s,m_1,m_2)\, q_{\mu} q_{\nu} ,
\label{E26b}
\ee
\esub
where, throughout this section, $s \equiv q^2$.

Threshold effects involving the $t\bar{t}$, $t\bar{b}$, and $b\bar{b}$
channels can be expressed as contributions to the imaginary parts of the
functions $\Pi^{V,A}$ and $\lambda^{V,A}$~\cite{R24,R27,R28,R29,R30,R31}.
A number of papers~\cite{R23,R24} have made use of DRs
to express the physically important amplitudes
$\Pi^{V,A}$ in terms of their imaginary parts.
In Refs.\cite{R30,R31} two of us (B.K.\ and A.S.) proposed to use DRs
directly constructed from the Ward identities.
We recall the basic strategy: contracting both sides of Eq.~(\ref{E26a})
with $q^{\mu}$, one derives the relation
\bsub
\be
 \Pi^{V,A}(s,m_1,m_2) = - s \lambda^{V,A}(s,m_1,m_2) +
 \Delta^{V,A}(s,m_1,m_2) ,
\label{E27a}
\ee
where $\Delta^{V,A}(s,m_1,m_2)$ is defined by
\be
  \int\,d^4x\,e^{i q\cdot{x}} \bra{0}T[\partial^{\mu}J_{\mu}^{V,A}(x)
   J_{\nu}^{V,A\,\dagger}(0)]\ket{0}   \equiv
   \Delta^{V,A}(s,m_1,m_2)\, q_{\nu} .
\label{E27b}
\ee
\esub
The idea then is to write DRs for $\lambda^{V,A}$ and
$\Delta^{V,A}$ and to obtain $\Pi^{V,A}$ by means of Eq.~(\ref{E27a}).
This approach has been employed in Refs.\cite{R30,R31} to discuss both
the perturbative and the threshold contributions in the on-shell scheme of
renormalization.
The aim of this section is to extend the discussion, so that the threshold
effects can also be included in the \ms\ calculations.

We recall that in our analysis the threshold contributions to the
imaginary parts are non-zero over a small, finite range, i.e.\  they
have compact support.
As a consequence, the corresponding unsubtracted DR
integrals for $\lambda^{V,A}$ and $\Delta^{V,A}$ are convergent and
moreover vanish as $|s| \rightarrow \infty$.
In using the DR approach, we self-consistently assume that the threshold
contributions to $\lambda^{V,A}$ and $\Delta^{V,A}$ tend to zero as $|s|
\rightarrow \infty$, so that these quantities satisfy unsubtracted DRs.
Thus,
\bsub
\bea
  \lambda^{V,A}(s,m_1,m_2) &=& \frac{1}{\pi} \int ds'\,
  	\frac{\Im \lambda^{V,A}(s',m_1,m_2)}{s'-s - i \epsilon} ,
	\label{E28a} \\*
  \Delta^{V,A}(s,m_1,m_2) &=& \frac{1}{\pi} \int ds'\,
        \frac{\Im \Delta^{V,A}(s',m_1,m_2)}{s'-s - i \epsilon} .
	\label{E28b}
\eea
\esub
Inserting these expressions into Eq.~(\ref{E27a}) and using the same
equation to relate the imaginary parts, one obtains the two equivalent
representations~\cite{R31}
\bsub
\bea
 \Pi^{V,A}(s,m_1,m_2) &=& \frac{1}{\pi} \int ds'
   \left[ \frac{\Im\Pi^{V,A}(s',m_1,m_2)}{s'-s - i \epsilon} +
      \Im\lambda^{V,A}(s',m_1,m_2) \right] ,
      \label{E29a} \\*
 \Pi^{V,A}(s,m_1,m_2) &=& \frac{s}{\pi} \int\frac{ds'}{s'}\,
       \frac{\Im\Pi^{V,A}(s',m_1,m_2)}{s'-s - i \epsilon}
       + \frac{1}{\pi} \int\frac{ds'}{s'}\Im\Delta^{V,A}(s',m_1,m_2) .
       \label{E29b}
\eea
\esub
In Eqs.~(\ref{E28a}--\ref{E29b}) and henceforth it is understood that
$\lambda^{V,A}$, $\Delta^{V,A}$, and $\Pi^{V,A}$ represent the excess
threshold contributions relative to the perturbative \Ord{\alpha \alpha_s}
corrections.
As explained in Ref.\cite{R31}, Eq.~(\ref{E29b}) can be directly derived
from the following assumptions:
{\it i}) $\Pi^{V,A}$ satisfies a once-subtracted DR;
{\it ii}) the subtraction constant is determined from the Ward identity
(\ref{E27a}), so that $\Pi^{V,A}(0) = \Delta^{V,A}(0)$;
{\it iii}) $\Delta^{V,A}(s)$ satisfies an unsubtracted DR, so that
$\Delta^{V,A}(0)$ can be calculated from the second integral in
Eq.~(\ref{E29b}).
The fact that $\Pi^{V,A}$ must satisfy a subtracted DR can be clearly
seen by considering the particular case of vector currents with equal
masses.
In that case $\lambda^{V}(s',m,m) = - \Pi^{V}(s',m,m)/s'$ and
$\Delta^{V}(s',m,m) = 0$, so that Eqs.~(\ref{E29a},b) reduce to
\bsub
\be
  \Pi^V(s,m,m) = \frac{s}{\pi} \int\frac{ds'}{s'}\,
  \frac{\Im\Pi^V(s',m,m)}{s' - s - i \epsilon} ,
\label{E30a}
\ee
which vanishes at $s = 0$, in conformity with the Ward identity (\ref{E27a}).
If, instead, $\Pi^V(s,m,m)$ were to satisfy an unsubtracted DR,
the condition $\Pi^V(0,m,m) = 0$ would imply \newline
$\int ds'\,\Im\Pi^V(s',m,m)/s' = 0$, which is manifestly false, as
$\Im\Pi^V(s',m,m) \geq 0$.
In summary, Eqs.~(\ref{E29a},b) are the simplest possible DRs consistent
with the Ward identity (\ref{E27a}).
As mentioned before, the latter is a crucial requirement.
We also see from Eq.~(\ref{E29a}) that the threshold effects modify
the asymptotic behavior of the full $\Pi^{V,A}(s)$ as $|s| \rightarrow
\infty$ by constants, i.e.\  by subleading terms. (We recall that the
perturbative contributions to $\Pi^{V,A}(s)$ grow as $s$, modulo
logarithms.)

Threshold effects associated with the $t\bar{b}$ channels are greatly
suppressed because they are proportional to the squared reduced mass of
the quarks~\cite{R24} and can be neglected.
Those involving the $b\bar{b}$ channel, i.e.\  ``bottomium
resonances,'' give significant contributions only to $e^2\pigg(0)$ and
are already included in the evaluation of $e^2\Re\left(\piggp{5}(0) -
\piggp{5}(\mzs)\right)$.
Thus, only the case $m_1 = m_2 = \mt$ is significant.
For vector currents with  equal masses, the relevant DR is given in
Eq.~(\ref{E30a}).
For axial-vector currents, we employ
\be
   \Pi^{A}(s,m,m) = \frac{1}{\pi} \int ds'
      \left[ \frac{\Im\Pi^A(s',m,m)}{s'-s - i \epsilon} +
            \Im\lambda^A(s',m,m) \right] ,
\label{E30b}
\ee
in accordance with Eq.~(\ref{E29a}).
The amplitude $\Im\Pi^A(s',\mt,\mt)$ receives contributions from
$J^P = 1^+$ states, i.e.\  $l = 1$,
which are suppressed near threshold by centrifugal barrier effects.
On the other hand, $\Im\lambda^A(s',\mt,\mt)$ also receives  significant
contributions from $0^-$ states, i.e.\  $l = 0$.
A detailed discussion of $\Im\Pi^V(s',\mt,\mt)$ and
$\Im\lambda^A(s',\mt,\mt)$ in both the resonance~\cite{R24} and
Green-function~\cite{R28} approaches is given in Ref.\cite{R31}.
In both cases one finds~\cite{R31}
\be
 \Im\lambda^A(s',\mt,\mt) \approx \Im\lambda^V(s',\mt,\mt) =
 -{\Im\Pi^V(s',\mt,\mt)\over s'} .
\label{E30c}
\ee
\esub
The second equality is, of course, an exact consequence of the Ward
identity (\ref{E27a}).

We now discuss the specific threshold contributions to the basic
radiative corrections studied in the paper.
The corrections $\Delta_\gamma$ and $\drcarw$ contain $e^2\piggp{f}(0)$
(cf.\ Eqs.~(\ref{E10a},b)).
The top contribution to this amplitude is obtained from Eq.~(\ref{EA7}),
\bsub
\be
 e^2\piggp{t}(0) = e^2 \frac{4}{9} (\Pi^V)'(0,\mt,\mt) ,
\label{E31a}
\ee
where $(\Pi^V)'(0,\mt,\mt) = (\partial/\partial s)\Pi^V(s,\mt,\mt)|_{s=0}$.
Recalling Eq.~(\ref{E30a}), we have
\be
 (\Pi^V)'(0,\mt,\mt) = \frac{1}{\pi} \int {ds'\over {s'}^2}
 \Im\Pi^V(s',\mt,\mt) .
\label{E31b}
\ee
\esub
There are no additional significant threshold effects in \drcarw,
because those involving
\newline $\Re\left(\Aww(\mws)-\Aww(0)\right)/\mws$
(cf.\ Eq.~(\ref{E10a})) are suppressed by reduced-mass effects.
In the case of \drcar\ (cf.\ Eq.~(\ref{E14})) the same holds true for the
term involving $\Re\Aww(\mws)/(\ccs\mzs)$; there are, however, significant
threshold contributions to $\Re\Azz(\mzs)/\mzs$.
According to Eq.~(\ref{EA13}),
\bsub
\be
 \frac{\Re\Azzp{(t)}(\mzs)}{\mzs} =
  -\frac{1}{16 \ccs \mzs}\Re\left[
   \left(1-\frac{8}{3}\scs\right)^2 \Pi^V(\mzs,\mt,\mt) +\Pi^A(\mzs,\mt,\mt)
   \right] .
\label{E32a}
\ee
Using Eqs.~(\ref{E30a}--c), we have
\bea
 \frac{\Re\Pi^V(\mzs,\mt,\mt)}{\mzs} &=&
   \frac{1}{\pi}\,{\cal P}\!\int\frac{ds'}{s'}\,
    \frac{\Im\Pi^V(s',\mt,\mt)}{s' - \mzs} ,
 \label{E32b} \\*
 \frac{\Re\Pi^A(\mzs,\mt,\mt)}{\mzs} &\approx&
  \frac{1}{\pi\mzs} \int ds'\, \Im\lambda^A(s',\mt,\mt)
  \nonumber\\*
 &\approx& -\frac{1}{\pi\mzs} \int\frac{ds'}{s'}\,
  \Im\Pi^V(s',\mt,\mt) ,
 \label{E32c}
\eea
\esub
where $\cal P$ denotes the principal value of the integral.
We note that, as explained earlier, the threshold behavior of
$\Im\Pi^A(s',\mt,\mt)$ is suppressed by centrifugal barrier effects and
its contribution has been neglected in Eq.~(\ref{E32c}).
As the support for the threshold contributions to $\Im\Pi^V(s',\mt,\mt)$
is located in the neighborhood of $s' \approx 4 \mts$, the dominant
effect for $\mts \gg \mzs$ is given by Eq.~(\ref{E32c}), with
Eqs.~(\ref{E31b},\ref{E32b}) being relatively suppressed by a factor
$\mzs/\left(4\mts\right)$.
Furthermore, the term involving $\Re\Pi^V(\mzs,\mt,\mt)$ in Eq.~(\ref{E32a})
has a small cofactor $\left(1-8\hat s^2/3\right)^2\approx0.14$.
In the range $\mz \leq \mt \leq 250 \GeV$, Eq.~(\ref{E32c}) increases
more rapidly than linearly with \mt, while the contributions of
Eqs.~(\ref{E31b},\ref{E32b}) to \drcar\ and \dr\ remain very small,
at most a few times $10^{-5}$.
As discussed in Ref.\cite{R31}, the sign of Eq.~(\ref{E32c}) can be
understood with an argument reminiscent of the one employed in
technicolor theories to explain the generation of the vector-boson
masses:
$0^-$ states contribute to the functions $\lambda^A(s)$ and
$\Delta^A(s)$ and, via the Ward identity (\ref{E27a}), to $\Pi^A(s)$.
In technicolor theories this gives rise to $\mzs \sim \Azz(0)$.
In our case, when Eq.~(\ref{E32c}) is inserted in Eq.~(\ref{E32a}), it
leads to a positive contribution to the \Z\ mass shift, $\delta\mzs =
\Re\Azz(\mzs)$.

Inserting into Eqs.~(\ref{E31b},\ref{E32b},c) the detailed evaluations of
$\Im\Pi^V(s',\mt,\mt)$ and
\newline $\Im\lambda^A(s',\mt,\mt)$~\cite{R31} derived
from the analyses of Refs.\cite{R24,R28}, we obtain the additional
contributions to \drcarw\ and \drcar\ associated with the $t\bar{t}$
threshold.
Because the threshold contributions in $e^2 \piggp{t}(0)$ and
$\Re\Agz(\mzs)/\mzs$ are very small, we have not subtracted them
from $\Delta_{\gamma}$ or the latter amplitude.
Consequently, they do not affect the counterterms of
Eqs.~(\ref{E7d},\ref{E9a}).

Analogous considerations are valid in the on-shell evaluation of \dr\
(see Section~5 and Ref.\cite{R31}).
The threshold corrections that are not suppressed by reduced-mass effects occur
again in $\piggp{t}(0)$ and $\Re\Azzp{(t)}(\mzs)/\mzs$, the latter being by
far the dominant contribution for large \mt.
Here one sets $\scs \rightarrow s^2$ and $\ccs \rightarrow c^2$ everywhere.
We also recall that in \dr\ the contribution of Eq.~(\ref{E32a}) is
enhanced by a factor $c^2/s^2$.

We end this section with the observation that the magnitude of the
1S contribution to $\Im\Pi^V(s,\mt,\mt)/s$ can
be roughly understood on the basis of a simplified ``Bohr-atom'' model.
We recall that, neglecting hard-gluon corrections, this contribution is
approximately given by~\cite{R24,R31}
\be
{\Im\Pi^V(s,\mt,\mt)\over s}\approx3\,{|R_{1,0}(0)|^2\over M_\theta}\,
\delta\left(s-M_\theta^2\right),
\label{E33}
\ee
where $R_{1,0}(0)$ is the radial wave function at the origin and
$M_\theta\approx2\mt$ is the mass of the 1S resonance, $\theta$.
In Ref.\cite{R24}, $|R_{1,0}(0)|^2$ and $M_\theta$ have been studied in
detail on the basis of the Richardson potential.
Suppose now that we attempt to estimate this effect using the ``Bohr-atom''
picture.
In the case of the one-electron atom,
$(R_{1,0}(0))^2=4(\alpha\mu)^3$, where $\mu$ is the reduced mass.
For toponium we set $\mu=\mt/2$, $M_\theta=2\mt$, and replace
$\alpha\rightarrow4\alphah_s(k_1)/3$, where
$k_1=2\alphah_s(k_1)\mt/3$ is the inverse Bohr radius~\cite{R27}.
This leads to
$3|R_{1,0}(0)|^2/\left(M_\theta\mzs\right)
=(16/9)\left(\alphah_s(k_1)\right)^3\mts/\mzs$, which is to be compared
with $3\mt\xi(\mt)/\left(8\mzs\right)$, obtained from the Richardson
potential~\cite{R24,R31}.
Here $\xi(\mt)$ is a monotonically increasing function of \mt, which
is evaluated numerically and varies from 1.95 to 3.08~GeV for
$\mz\le\mt\le250\GeV$.
For $\mt= \mz$, 250~GeV,
the Richardson potential gives 0.0080 and 0.0347, respectively,
while in the ``Bohr-atom'' model the corresponding values are
0.0096 and 0.0428, which are about 20\% larger.
This may be partly due to the fact that the Richardson potential is softer than
Coulombic near the origin.
Interestingly, the ratio of the values at $\mt=250\GeV$ and $\mt=\mz$,
which gives an indication of the \mt\ dependence, is almost the same in both
cases.
In the ``Bohr-atom'' model, for the $n$S states, $|R_{n,0}(0)|^2$ scales as
$n^{-3}$.
However, as the radii of the higher orbits increase as $n^2$, the momentum
$k_n$ at which $\alphah_s$ is to be evaluated becomes smaller.
For sufficiently high $n$, the relevant $\alphah_s$ falls in the
non-perturbative regime and the ``Coulombic'' picture becomes increasingly
doubtful.
The evaluation  based on a realistic, ``confining'' potential is clearly
preferable.
However, it is interesting that a simple Coulomb potential gives a
similar answer, at least in the case of the 1S state.

\section{Numerical Calculations}

  In the previous sections we have discussed the theoretical background
necessary for the incorporation of the leading QCD effects in the basic
corrections \drcarw, \drcar, and \dr.
At the same time, as explained in Section~1, we have introduced a number
of refinements in the analysis of the electroweak corrections.

In this section, we apply the previous results to numerically evaluate
\drcarw, \drcar, and \dr, and, most importantly, to derive precise values for
\mw\ and $\scs \equiv \sincarmz$, as functions of \mt\ and \mh.
Working first in the \ms\ scheme, the basic strategy is the same as in
Ref.\cite{R1}.
We employ Eqs.~(\ref{E10c}) and (\ref{E14},\ref{E18a}) as the basic
expressions for \drcarw\ and \drcar, respectively, and iteratively
evaluate these corrections in conjunction with Eq.~(\ref{E19a}), leading
to accurate values for \scs.
Then \dr\ can be evaluated from Eqs.~(\ref{E19c}) or (\ref{E19e}), $s^2$ from
Eq.~(\ref{E19b}), and \mw\ from either $s^2$ or Eq.~(\ref{E19f}).
We use as input values $\alpha=(137.0359895)^{-1}$,
$\Gmu=1.16639\times10^{-5}\GeV^{-2}$~\cite{R50}, $\mz=91.187\GeV$~\cite{R16},
leading to $A = (\pi \alpha/\sqrt{2}\Gmu)^{1/2} = 37.2802\GeV$,
and $\alphah_s(\mz) = 0.118$~\cite{R38}.
As shown in Appendix~C, the effects arising from finite fermion masses
are very minor.
Nonetheless, we include them as follows: the $u$, $d$, and $s$ quarks are
treated as massless, while we employ $m_c = 1.5\GeV$ and $m_b = 4.5\GeV$.
The leptons are given their physical masses~\cite{R50},
including the recent value $m_{\tau} = 1.777\GeV$~\cite{R51}.
In order to incorporate the new results on the leading irreducible
corrections of \Ord{\alpha^2}~\cite{R7}, it is very convenient to use a
precise analytical representation of the function $R(\mh/\mt)$ in
Eqs.~(\ref{E17a},\ref{E18a},\ref{E23b}).
In the range $r=\mh/\mt > 4$, the authors of Ref.\cite{R7} give the
accurate asymptotic expansion
\bsub
\begin{eqnarray}
 R(r) &=& \frac{49}{4}+\pi^2 - 27 \ln r + 6 \ln^2r
 \nonumber \\
      & & + \frac{1}{3\,r^2} (2 - 12 \pi^2 - 24 \ln{r} - 108 \ln^2r)
  \nonumber \\
      & & + \frac{1}{48\,r^4} (1613 - 240 \pi^2 + 3000 \ln r - 2880 \ln^2r).
\label{E34a}
\end{eqnarray}
In the complementary domain $0 < r \leq 4$, only numerical values are
available~\cite{R7}, which we have fitted with the expression
\begin{eqnarray}
  R(r) &=& -0.7392088 + r (-11.5315 + 0.382497 \ln{r})
  \nonumber \\
       & & +\,r^2 (5.31338 - 3.055 \ln{r} + 0.523039 \ln^2r) .
\label{E34b}
\end{eqnarray}
\esub
The maximum deviation of Eq.~(\ref{E34b}) from the original data~\cite{R7}
is $\lequiv\, 0.025$ and occurs near the matching point, $r=4$.

In Tables I--V we display the calculated values of \mw\ and \scs, as
functions of \mt\ for $\mh = (60$, 100, 250, 600, 1000)~GeV.
Currently, a fit to all data, using the electroweak radiative corrections
of the SM and $\alpha_s(\mz)=0.120\pm0.006$, gives
$\mt=145{+17+17\atop-19-19}\GeV$~\cite{R51}.
The central value corresponds to $\mh=300\GeV$ and the last error reflects
the theoretical uncertainty associated with the range
$60\GeV\le\mh\le1\TeV$.
This implies $\mt\le173{+17\atop-19}\GeV$ at the 95\% confidence level,
where the error is again due to the \mh\ uncertainty.
Although this strongly suggests that $\mt\,\lequiv\,200\GeV$, in the tables
we present values up to $\mt=250\GeV$.
One of the reasons is that it is theoretically interesting to study and
compare the high-\mt\ dependence of the various radiative corrections.
Moreover, there is always the possibility of a statistical surprise or
that unknown new physics may alter the predictions of the SM, so that
scenarios in which \mt\ is found at higher values are not completely
excluded.
For comparison purposes, we list in the tables the results of four different
calculations:
{\it i}) only electroweak corrections, with all QCD corrections turned
off (columns labelled EW for ``electroweak'');
{\it ii}) electroweak plus perturbative \Ord{\alpha \alpha_s}
corrections (columns labelled P for ``perturbative'');
{\it iii}) the above, plus threshold effects calculated in the resonance
approach according to Section~6 (columns labelled P+R for ``perturbative
plus resonance'');
{\it iv}) the same as in ({\it iii}) but with threshold effects evaluated
in the Green-function approach (columns labelled P+G for ``perturbative
plus Green function'').
In order to keep the tables compact, we have displayed only the
quantities of greatest physical interest, namely \mw\ and \scs, rather
than the radiative corrections \drcarw, \drcar, and \dr, or the derived
parameter $s^2$ of the on-shell scheme.
The interested reader can readily glean these important quantities from
the tables.
Thus, inserting the value of \scs\ in Eq.~(\ref{E3}) and those of \scs\
and \mw\ in Eq.~(\ref{E2}), one finds \drcar\ and \drcarw, respectively.
Using \mw, one calculates $s^2 \equiv 1 - \mws/\mzs$ and, in conjunction
with Eq.~(\ref{E1}), \dr.
We have kept enough decimal figures in \scs\ and \mw, so that
\drcarw, \drcar, \dr, and $s^2$ can be accurately evaluated.
The tables allow us to separate the threshold effects from the more
established perturbative \Ord{\alpha \alpha_s} contributions.
Although the resonance (R) and Green-function (G) approaches are quite
different conceptually and technically, the tables reveal the welcome
and rather surprising result that their effect on \mw\ and \scs\ is
very similar over the entire range $\mz \leq \mt \leq 250\GeV$,
$60\GeV \leq \mh \leq 1 \TeV$.
Nonetheless, for reasons explained in Ref.\cite{R31} and Section~6, in
our specific applications we use the resonance method for $\mt\, \lequiv\, 130
\GeV$ and the Green-function approach for $\mt \gequiv 130\GeV$.

It is also a curious and rather surprising fact that most of the new
effects we have considered relative to Ref.\cite{R1}, namely
{\it a}) the incorporation of the recent results on the leading
irreducible corrections of \Ord{\alpha^2},
{\it b}) the perturbative \Ord{\alpha \alpha_s} corrections, and
{\it c}) the threshold contributions in the formulation of Section~6,
increase the values of \drcar\ and \dr\ for given \mt\ and \mh.
Thus, they have a sign opposite to that of the leading \mt-dependent
part of the one-loop corrections and, therefore, they induce an
increase in the \mt\ upper bound.
For large \mt, they are only partially compensated by the shift of
$-5\times10^{-4}$ in \drcar\ and \dr\ arising from the new
calculation of $e^2\Re\left(\piggp{5}(0) - \piggp{5}(\mzs)\right)$~\cite{R10}.
It should also be remembered that most of the corresponding shifts in
\drcar\ and \dr\ increase with \mt:
({\it a}) behaves as $m_t^4$ (cf.\ Eqs.~(\ref{E17a},\ref{E18a},\ref{E23b})),
the dominant ({\it b}) contributions as \mts\ (cf.\
Eqs.~(\ref{E16a},\ref{E19c})), and
({\it c}) more rapidly than linearly in \mt~\cite{R31}.

The QCD effects on \mw\ and \scs\ are visible in the tables.
For example, for the intermediate value $\mh = 250\GeV$ and
$\mt = \mz$, 130, 150, 180, 200, 250~GeV, the perturbative
\Ord{\alpha \alpha_s} corrections lead to the shifts
\bsub
\be
\left.
  \begin{array}{ccc}
  \Delta\mw & = & -(32,42,50,65,77,113)\MeV\\
  \Delta\scs & = & +(1.4,2.3,2.8,3.7,4.4,6.4)\times10^{-4}
  \end{array}
\right\}\, \Ord{\alpha \alpha_s}.
\label{E35a}
\ee
The threshold contributions lead to further shifts
\be
\left.
  \begin{array}{ccc}
  \Delta\mw & = & -(8, 14, 14, 20, 25, 42)\MeV\\
  \Delta\scs & = & +(0.7, 0.9, 0.9, 1.2, 1.5, 2.5)\times10^{-4}
  \end{array}
\right\}\, \mbox{(threshold)}.
\label{E35b}
\ee
\esub
The departure from the monotonic behaviour between $\mt = 130\GeV$ and
$\mt = 150\GeV$ is due to the transition from the resonance to the
Green-function approach.
The effect of the leading irreducible corrections of
\Ord{(\alpha\,\mts/\mws)^2} cannot be seen in the tables because they have been
included in every column.
However, they can be estimated from the relations
\bsub
\bea
 \frac{\Delta\mw}{\mw} &\approx& \frac{c^2}{c^2-s^2-2c^2x_t}\,
 R\left({\mh\over\mt}\right)\frac{x_t^2}{6} ,
 \label{E36a} \\*
 \Delta\scs &\approx& -\frac{\scs \ccs}{\ccs - \scs}\,
 R\left({\mh\over\mt}\right)\frac{x_t^2}{3(1-x_t)^2} ,
 \label{E36b}
\eea
where $x_t$ is defined in Eq.~(\ref{E17b}).
The term $2c^2x_t$ in the first denominator of Eq.~(\ref{E36a})
and the factor $(1-x_t)^{-2}$ in Eq.~(\ref{E36b}) take into account the
fact that some of the leading one-loop contributions to \dr\ and \drcar\
depend on $s^2$ and \scs, respectively, and are therefore affected by the
shifts in these parameters.
For the same values of \mh\ and \mt\ as employed before, Eqs.~(\ref{E36a},b)
give
\be
\left.
  \begin{array}{ccc}
  \Delta\mw & = & -(1, 5, 8, 16, 24, 52) \MeV\\
  \Delta\scs & = & +(0.1, 0.3, 0.5, 0.9, 1.4, 3.0)\times10^{-4}
  \end{array}
\right\}\, \Ord{\alpha^2} .
\label{E36c}
\ee
\esub
Equations~(\ref{E35a},b,\ref{E36c}) can be compared with the
experimental uncertainties $(\delta\mw)_{\rm exp} \approx 100 \MeV$ and
$(\delta\scs)_{\rm exp} \approx 4\times10^{-4}$ expected at the end of
1993~\cite{R51}.
We see that, in general terms, they are of the same order of magnitude.
Of course, in the long run, even better experimental accuracies are
envisaged, reaching perhaps $(\delta\mw)_{\rm exp}\approx50\MeV$.
The above shifts can also  be compared with the theoretical uncertainties
\bsub
\bea
 (\delta\mw)_{\rm th} &=& - {\mw\over2}\,\frac{s^2}{c^2-s^2-2c^2x_t}\,
          \frac{\delta\dr}{1-\Delta\alpha} \approx \pm 16 \MeV ,
 \label{E37a} \\*
 (\delta\scs)_{\rm th} &=& \frac{\ccs \scs}{(\ccs - \scs)}\,
 \frac{\delta\drcar}{1 - \Delta_\gamma} \approx \pm 3\times10^{-4} ,
 \label{E37b}
\eea
\esub
arising from the error $(\delta\dr)_{\rm th} \approx
(\delta\drcar)_{\rm th} = \pm 9\times10^{-4}$ in the calculation of
$e^2\Re\left(\piggp{5}(0)\right.$ $\left.-\piggp{5}(\mzs)\right)$~\cite{R10}.
We recall that the theoretical error arising from the neglect of higher-order
electroweak corrections is expected to be of \Ord{(\alpha/\pi s^2)x_t}
in \drcarw\ and \drcar\ and further enhanced by a factor $c^2/s^2$
in \dr\ (see Section~1).
Moreover, the error in the overall QCD corrections (perturbative
and threshold effects) has been estimated to be $\pm20\%$~\cite{R31}.
Incidentally, Eq.~(\ref{E37b}) shows that, if the experimental accuracy in
\scs\ is improved in the future well beyond $4\times10^{-4}$, a meaningful
theoretical interpretation will require a decrease in the
above-mentioned theoretical errors.

As pointed out before, the higher-order corrections we have considered
lead to an increase in the \mt\ values obtained from experiments.
As an example, we consider the case $\mh = 250\GeV$ and call \mt\ the
parameter derived from $(\mw)_{\rm exp}$ when the perturbative
\Ord{\alpha \alpha_s} corrections are included.
Table~III shows that the possible values $\mt = \mz$, 130, 150, 180, 200~GeV
are larger than those obtained in the purely EW calculation by
\bsub
\be
 \Delta\mt = +(5.8, 7.5, 8.3, 9.7, 10.8)\GeV\quad(\Ord{\alpha\alpha_s}).
\label{E38a}
\ee
The values of \mt\ derived from $(\scs)_{\rm exp}$ are shifted by slightly
higher amounts,
the differences with Eq.(\ref{E38a}) being $\lequiv\, 0.8\GeV$.
The bulk of
$\Delta\mt$ arising from the perturbative \Ord{\alpha \alpha_s}
corrections can be understood with the simple formula (\ref{E16c}),
which describes the dominant contribution.
Similarly, we see  from Table~III that, when threshold contributions are
included, the values of \mt\ derived from $(\mw)_{\rm exp}$ are larger than
those obtained in the EW+P calculation by additional shifts of
\be
 \Delta\mt = +(1.7, 2.7,2.7, 3.3, 4.1) \GeV\quad(\mbox{threshold}).
\label{E38b}
\ee
The corresponding variations arising from the irreducible \Ord{\alpha^2}
corrections are
\be
 \Delta\mt = +(0.6, 1.2, 1.6, 2.5, 3.3)\GeV\quad(\Ord{\alpha^2}).
\label{E38c}
\ee
\esub
Thus, for $\mh = 250\GeV$ and $\mt = 200\GeV$, the combination of
\Ord{\alpha \alpha_s}, threshold, and leading irreducible \Ord{\alpha^2}
corrections increases the value of \mt\ derived from $(\mw)_{\rm exp}$ by
$\approx 16.8\GeV$.

Another topic of considerable interest is the comparison of calculations
carried out in the \ms\ and on-shell methods of renormalization.
This is illustrated for $\mh = 60$, 250, 1000~GeV in Tables~VI--VIII,
where we show the evaluation of \dr, a physical observable, obtained on the
basis of the \ms\ approach of Section~3 and the on-shell formulation of
Section~5.
For the purposes of this study, we have included the electroweak and
perturbative \Ord{\alpha \alpha_s} corrections, leaving aside the
threshold effects.
We also display the derived values of \mw\ and $s^2$.
Inspection of the tables shows that the two calculations of \dr\ are in
excellent agreement over the entire range $60\GeV \leq \mh \leq 1 \TeV$,
$\mz \leq \mt \leq 250 \GeV$, with a maximum discrepancy of $\approx
2.5\times10^{-4}$ occurring at $\mt = 250 \GeV$ and $\mh = 1 \TeV$.
Although such accurate agreement may be somewhat fortuitous, it is
roughly of the expected order of magnitude, i.e.\
\Ord{(\alpha/\pi\scs)(\ccs/\scs) x_t}.

\section*{Acknowledgments}
We would like to thank Riccardo Barbieri, Matteo Beccaria, Giuseppe
Degrassi, Fred Jegerlehner, Hans K\"uhn,
William Marciano, Alfred Mueller, Lev Okun, Michael Peskin,
Thomas Teubner, and Peter Zerwas for valuable communications and
discussions.
One of us (A.S.) would like to thank the physicists at SISSA, Trieste,
Italy, for their kind hospitality during June~1992, when part of his
work on this problem was carried out.
This research was supported in part by the National Science Foundation
under Grant No.~PHY-9017585. One of the authors (S.F.) is currently
supported by a ICSC World Laboratory Fellowship
and the CERN Theory Division.

\renewcommand{\arraystretch}{1.2}
\begin{table}[h]
\centering
\begin{tabular}{|c|c|ccc|c|ccc|}
  \hline
     \mt
     &  \multicolumn{4}{c|}{$\hat{s}^2$ }
     &  \multicolumn{4}{c|}{$ \mw \mbox{\small [GeV]} $} 		   \\
  \cline{2-9}
     \mbox{\small[GeV]} & \lower 1.5ex\hbox{EW}
     & \multicolumn{3}{c|}{EW + QCD}
     & \lower 1.5ex\hbox{EW} & \multicolumn{3}{c|}{EW + QCD} \\
  \cline{3-5} \cline{7-9} 
   {} & {} & P & P+R & P+G & {} & P  & P+R & P+G  \\
        \hline
\mz        &   0.23307 &   0.23321 &   0.23327 &   0.23326 &   79.979
 &   79.947 &   79.938 &   79.941 \\
100      &   0.23287 &   0.23303 &   0.23309 &   0.23307 &   80.029 &   79.996
 &   79.986 &   79.989 \\
110      &   0.23262 &   0.23280 &   0.23287 &   0.23285 &   80.085 &   80.050
 &   80.038 &   80.041 \\
120      &   0.23235 &   0.23255 &   0.23263 &   0.23261 &   80.141 &   80.103
 &   80.091 &   80.094 \\
130      &   0.23207 &   0.23229 &   0.23238 &   0.23236 &   80.200 &   80.158
 &   80.144 &   80.148 \\
140      &   0.23176 &   0.23201 &   0.23211 &   0.23209 &   80.260 &   80.215
 &   80.199 &   80.203 \\
150      &   0.23144 &   0.23172 &   0.23182 &   0.23180 &   80.323 &   80.274
 &   80.256 &   80.260 \\
160      &   0.23110 &   0.23140 &   0.23152 &   0.23150 &   80.389 &   80.335
 &   80.316 &   80.319 \\
170      &   0.23074 &   0.23108 &   0.23120 &   0.23118 &   80.457 &   80.398
 &   80.377 &   80.380 \\
180      &   0.23036 &   0.23073 &   0.23087 &   0.23085 &   80.528 &   80.463
 &   80.441 &   80.444 \\
190      &   0.22997 &   0.23037 &   0.23052 &   0.23050 &   80.603 &   80.532
 &   80.507 &   80.510 \\
200      &   0.22955 &   0.22999 &   0.23015 &   0.23013 &   80.680 &   80.603
 &   80.576 &   80.578 \\
210      &   0.22912 &   0.22959 &   0.22976 &   0.22976 &   80.760 &   80.677
 &   80.648 &   80.650 \\
220      &   0.22867 &   0.22918 &   0.22936 &   0.22936 &   80.844 &   80.754
 &   80.723 &   80.723 \\
230      &   0.22820 &   0.22875 &   0.22895 &   0.22895 &   80.930 &   80.833
 &   80.800 &   80.799 \\
240      &   0.22772 &   0.22831 &   0.22851 &   0.22853 &   81.020 &   80.916
 &   80.881 &   80.878 \\
250      &   0.22721 &   0.22785 &   0.22806 &   0.22809 &   81.114 &   81.001
 &   80.964 &   80.959 \\
  \hline
\end{tabular}
\caption[TableI]{
Calculated values of \sincarmz\ and \mw, as a function of \mt, for $\mz
= 91.187 \GeV$, $\alphah_s(\mz) = 0.118$, and $\mh = 60 \GeV$.
The EW column includes only electroweak radiative corrections.
The P column incorporates perturbative \Ord{\alpha\alpha_s}
contributions (see Section~3). The P+R and P+G columns contain also  $t
\bar{t}$ threshold effects (see Section~6) in the resonance and
Green-function approaches, respectively.
The on-shell parameter $\sin^2\theta_{\scriptscriptstyle W}$ and the
radiative corrections \drcarw, \drcar, and \dr\ can be gleaned from this
table, as explained in Section~7.}
\end{table}

\begin{table}[h]
\centering
\begin{tabular}{|c|c|ccc|c|ccc|}
  \hline
     \mt
     &  \multicolumn{4}{c|}{$\hat{s}^2$ }
     &  \multicolumn{4}{c|}{$ \mw \mbox{\small [GeV]} $} 		   \\
  \cline{2-9}
     \mbox{\small[GeV]} & \lower 1.5ex\hbox{EW}
     & \multicolumn{3}{c|}{EW + QCD}
     & \lower 1.5ex\hbox{EW} & \multicolumn{3}{c|}{EW + QCD} \\
  \cline{3-5} \cline{7-9} 
   {} & {} & P & P+R & P+G & {} & P  & P+R & P+G  \\
        \hline
 \mz        &   0.23332 &   0.23346 &   0.23352 &   0.23351 &   79.952
 &   79.921 &   79.912 &   79.914 \\
100      &   0.23312 &   0.23328 &   0.23334 &   0.23332 &   80.002 &   79.969
 &   79.960 &   79.962 \\
110      &   0.23287 &   0.23305 &   0.23312 &   0.23310 &   80.058 &   80.023
 &   80.012 &   80.015 \\
120      &   0.23260 &   0.23280 &   0.23289 &   0.23287 &   80.115 &   80.077
 &   80.064 &   80.068 \\
130      &   0.23232 &   0.23254 &   0.23263 &   0.23261 &   80.173 &   80.132
 &   80.118 &   80.121 \\
140      &   0.23201 &   0.23227 &   0.23236 &   0.23234 &   80.234 &   80.188
 &   80.172 &   80.176 \\
150      &   0.23169 &   0.23197 &   0.23208 &   0.23206 &   80.296 &   80.246
 &   80.229 &   80.233 \\
160      &   0.23135 &   0.23166 &   0.23178 &   0.23175 &   80.361 &   80.307
 &   80.288 &   80.292 \\
170      &   0.23100 &   0.23133 &   0.23146 &   0.23144 &   80.429 &   80.370
 &   80.349 &   80.352 \\
180      &   0.23062 &   0.23099 &   0.23113 &   0.23111 &   80.500 &   80.435
 &   80.412 &   80.416 \\
190      &   0.23023 &   0.23063 &   0.23078 &   0.23076 &   80.574 &   80.503
 &   80.478 &   80.481 \\
200      &   0.22982 &   0.23025 &   0.23041 &   0.23040 &   80.651 &   80.574
 &   80.547 &   80.549 \\
210      &   0.22939 &   0.22986 &   0.23003 &   0.23002 &   80.730 &   80.647
 &   80.618 &   80.619 \\
220      &   0.22894 &   0.22945 &   0.22964 &   0.22963 &   80.813 &   80.723
 &   80.692 &   80.692 \\
230      &   0.22848 &   0.22903 &   0.22922 &   0.22923 &   80.899 &   80.801
 &   80.769 &   80.767 \\
240      &   0.22800 &   0.22859 &   0.22879 &   0.22881 &   80.988 &   80.883
 &   80.848 &   80.845 \\
250      &   0.22750 &   0.22814 &   0.22835 &   0.22838 &   81.080 &   80.967
 &   80.930 &   80.925 \\
  \hline
\end{tabular}
\caption{As Table~I, for $\mh = 100 \GeV$.}
\end{table}

\begin{table}[h]
\centering
\begin{tabular}{|c|c|ccc|c|ccc|}
  \hline
     \mt
     &  \multicolumn{4}{c|}{$\hat{s}^2$ }
     &  \multicolumn{4}{c|}{$ \mw \mbox{\small [GeV]} $} 		   \\
  \cline{2-9}
     \mbox{\small[GeV]} & \lower 1.5ex\hbox{EW}
     & \multicolumn{3}{c|}{EW + QCD}
     & \lower 1.5ex\hbox{EW} & \multicolumn{3}{c|}{EW + QCD} \\
  \cline{3-5} \cline{7-9} 
   {} & {} & P & P+R & P+G & {} & P  & P+R & P+G  \\
        \hline
\mz        &   0.23380 &   0.23394 &   0.23401 &   0.23399 &   79.894
 &   79.862 &   79.854 &   79.856 \\
100      &   0.23360 &   0.23376 &   0.23383 &   0.23381 &   79.944 &   79.911
 &   79.901 &   79.904 \\
110      &   0.23335 &   0.23353 &   0.23361 &   0.23359 &   80.000 &   79.965
 &   79.953 &   79.956 \\
120      &   0.23308 &   0.23329 &   0.23337 &   0.23335 &   80.056 &   80.018
 &   80.005 &   80.009 \\
130      &   0.23280 &   0.23303 &   0.23312 &   0.23310 &   80.114 &   80.072
 &   80.058 &   80.062 \\
140      &   0.23250 &   0.23275 &   0.23285 &   0.23283 &   80.174 &   80.128
 &   80.113 &   80.116 \\
150      &   0.23218 &   0.23246 &   0.23257 &   0.23255 &   80.236 &   80.186
 &   80.169 &   80.172 \\
160      &   0.23185 &   0.23215 &   0.23227 &   0.23225 &   80.301 &   80.246
 &   80.227 &   80.230 \\
170      &   0.23149 &   0.23183 &   0.23196 &   0.23194 &   80.368 &   80.308
 &   80.287 &   80.290 \\
180      &   0.23112 &   0.23149 &   0.23163 &   0.23161 &   80.437 &   80.372
 &   80.349 &   80.352 \\
190      &   0.23073 &   0.23114 &   0.23129 &   0.23127 &   80.510 &   80.439
 &   80.414 &   80.417 \\
200      &   0.23033 &   0.23077 &   0.23093 &   0.23092 &   80.585 &   80.508
 &   80.481 &   80.483 \\
210      &   0.22991 &   0.23039 &   0.23056 &   0.23055 &   80.663 &   80.579
 &   80.550 &   80.552 \\
220      &   0.22947 &   0.22999 &   0.23017 &   0.23017 &   80.744 &   80.653
 &   80.622 &   80.622 \\
230      &   0.22902 &   0.22958 &   0.22977 &   0.22978 &   80.827 &   80.730
 &   80.697 &   80.695 \\
240      &   0.22855 &   0.22915 &   0.22936 &   0.22937 &   80.914 &   80.809
 &   80.773 &   80.771 \\
250      &   0.22807 &   0.22871 &   0.22893 &   0.22896 &   81.003 &   80.890
 &   80.853 &   80.848 \\
  \hline
\end{tabular}
\caption{As  Table~I, for $\mh = 250 \GeV$.}
\end{table}

\begin{table}[h]
\centering
\begin{tabular}{|c|c|ccc|c|ccc|}
  \hline
     \mt
     &  \multicolumn{4}{c|}{$\hat{s}^2$ }
     &  \multicolumn{4}{c|}{$ \mw \mbox{\small [GeV]} $} 		   \\
  \cline{2-9}
     \mbox{\small[GeV]} & \lower 1.5ex\hbox{EW}
     & \multicolumn{3}{c|}{EW + QCD}
     & \lower 1.5ex\hbox{EW} & \multicolumn{3}{c|}{EW + QCD} \\
  \cline{3-5} \cline{7-9} 
   {} & {} & P & P+R & P+G & {} & P  & P+R & P+G  \\
        \hline
\mz        &   0.23430 &   0.23444 &   0.23450 &   0.23448 &   79.828
 &   79.796 &   79.787 &   79.790 \\
100      &   0.23409 &   0.23425 &   0.23432 &   0.23430 &   79.878 &   79.844
 &   79.835 &   79.837 \\
110      &   0.23384 &   0.23403 &   0.23410 &   0.23408 &   79.933 &   79.898
 &   79.887 &   79.890 \\
120      &   0.23358 &   0.23378 &   0.23386 &   0.23384 &   79.990 &   79.951
 &   79.939 &   79.942 \\
130      &   0.23329 &   0.23352 &   0.23361 &   0.23359 &   80.047 &   80.005
 &   79.991 &   79.995 \\
140      &   0.23299 &   0.23325 &   0.23335 &   0.23333 &   80.107 &   80.061
 &   80.045 &   80.049 \\
150      &   0.23268 &   0.23296 &   0.23307 &   0.23305 &   80.168 &   80.118
 &   80.101 &   80.104 \\
160      &   0.23234 &   0.23265 &   0.23277 &   0.23275 &   80.232 &   80.177
 &   80.158 &   80.162 \\
170      &   0.23200 &   0.23234 &   0.23247 &   0.23244 &   80.298 &   80.238
 &   80.217 &   80.221 \\
180      &   0.23163 &   0.23200 &   0.23214 &   0.23212 &   80.367 &   80.301
 &   80.278 &   80.282 \\
190      &   0.23125 &   0.23166 &   0.23181 &   0.23179 &   80.438 &   80.367
 &   80.342 &   80.344 \\
200      &   0.23086 &   0.23130 &   0.23146 &   0.23145 &   80.511 &   80.434
 &   80.407 &   80.409 \\
210      &   0.23045 &   0.23093 &   0.23110 &   0.23109 &   80.587 &   80.503
 &   80.475 &   80.476 \\
220      &   0.23002 &   0.23054 &   0.23072 &   0.23072 &   80.666 &   80.575
 &   80.544 &   80.544 \\
230      &   0.22958 &   0.23014 &   0.23034 &   0.23034 &   80.747 &   80.649
 &   80.616 &   80.615 \\
240      &   0.22913 &   0.22973 &   0.22994 &   0.22995 &   80.830 &   80.725
 &   80.689 &   80.687 \\
250      &   0.22867 &   0.22931 &   0.22953 &   0.22956 &   80.916 &   80.803
 &   80.765 &   80.760 \\
  \hline
\end{tabular}
\caption{As  Table~I, for $\mh = 600 \GeV$.}
\end{table}

\begin{table}[h]
\centering
\begin{tabular}{|c|c|ccc|c|ccc|}
  \hline
     \mt
     &  \multicolumn{4}{c|}{$\hat{s}^2$ }
     &  \multicolumn{4}{c|}{$ \mw \mbox{\small [GeV]} $} 		   \\
  \cline{2-9}
     \mbox{\small[GeV]} & \lower 1.5ex\hbox{EW}
     & \multicolumn{3}{c|}{EW + QCD}
     & \lower 1.5ex\hbox{EW} & \multicolumn{3}{c|}{EW + QCD} \\
  \cline{3-5} \cline{7-9} 
   {} & {} & P & P+R & P+G & {} & P  & P+R & P+G  \\
        \hline
\mz        &   0.23460 &   0.23474 &   0.23480 &   0.23479 &   79.786
 &   79.754 &   79.745 &   79.748 \\
100      &   0.23439 &   0.23455 &   0.23462 &   0.23460 &   79.836 &   79.803
 &   79.793 &   79.796 \\
110      &   0.23414 &   0.23432 &   0.23440 &   0.23438 &   79.892 &   79.856
 &   79.845 &   79.848 \\
120      &   0.23388 &   0.23408 &   0.23416 &   0.23414 &   79.948 &   79.910
 &   79.897 &   79.900 \\
130      &   0.23359 &   0.23382 &   0.23391 &   0.23389 &   80.006 &   79.964
 &   79.950 &   79.953 \\
140      &   0.23329 &   0.23355 &   0.23365 &   0.23362 &   80.065 &   80.019
 &   80.004 &   80.007 \\
150      &   0.23297 &   0.23326 &   0.23337 &   0.23334 &   80.127 &   80.077
 &   80.059 &   80.063 \\
160      &   0.23264 &   0.23295 &   0.23307 &   0.23305 &   80.190 &   80.136
 &   80.116 &   80.120 \\
170      &   0.23229 &   0.23264 &   0.23276 &   0.23274 &   80.256 &   80.196
 &   80.175 &   80.179 \\
180      &   0.23193 &   0.23230 &   0.23244 &   0.23242 &   80.325 &   80.259
 &   80.236 &   80.239 \\
190      &   0.23155 &   0.23196 &   0.23211 &   0.23209 &   80.395 &   80.324
 &   80.299 &   80.302 \\
200      &   0.23116 &   0.23160 &   0.23176 &   0.23175 &   80.468 &   80.390
 &   80.363 &   80.366 \\
210      &   0.23075 &   0.23123 &   0.23141 &   0.23140 &   80.543 &   80.459
 &   80.430 &   80.431 \\
220      &   0.23033 &   0.23085 &   0.23104 &   0.23104 &   80.621 &   80.530
 &   80.499 &   80.499 \\
230      &   0.22990 &   0.23046 &   0.23066 &   0.23066 &   80.700 &   80.602
 &   80.569 &   80.568 \\
240      &   0.22946 &   0.23006 &   0.23027 &   0.23028 &   80.782 &   80.677
 &   80.641 &   80.638 \\
250      &   0.22900 &   0.22965 &   0.22987 &   0.22990 &   80.866 &   80.753
 &   80.715 &   80.710 \\
  \hline
\end{tabular}
\caption{As  Table~I, for $\mh = 1000 \GeV$. }
\end{table}

\begin{table}[h]
\begin{center}
\begin{tabular}{|c|cc|cc|}
 \hline
  \mt    &   \drI      &   \drII     &  \mwI     &  \mwII   \\
 $\mbox{\small[GeV]}$ & \multicolumn{2}{c|}{$\times10^{2}$} &
 \multicolumn{2}{c|}{$\mbox{\small[GeV]}$}  \\
 \hline
 \mz    &      6.003  &      6.006  &   79.947  &   79.946  \\
 100  &      5.735  &      5.738  &   79.996  &   79.995  \\
 110  &      5.437  &      5.440  &   80.050  &   80.049  \\
 120  &      5.135  &      5.138  &   80.103  &   80.103  \\
 130  &      4.823  &      4.826  &   80.158  &   80.158  \\
 140  &      4.498  &      4.502  &   80.215  &   80.214  \\
 150  &      4.159  &      4.162  &   80.274  &   80.273  \\
 160  &      3.801  &      3.805  &   80.334  &   80.334  \\
 170  &      3.425  &      3.429  &   80.398  &   80.397  \\
 180  &      3.029  &      3.033  &   80.463  &   80.463  \\
 190  &      2.611  &      2.615  &   80.532  &   80.531  \\
 200  &      2.170  &      2.173  &   80.603  &   80.602  \\
 210  &      1.705  &      1.708  &   80.677  &   80.676  \\
 220  &      1.214  &      1.217  &   80.754  &   80.753  \\
 230  &      0.697  &      0.699  &   80.833  &   80.833  \\
 240  &      0.151  &      0.153  &   80.916  &   80.915  \\
 250  &     -0.424  &     -0.423  &   81.001  &   81.001  \\
 \hline
\end{tabular}
\caption[TableVI]{
 Comparison between the values of \dr\ and \mw\ obtained using the \ms\
approach
 of Section~3 (\drI, \mwI) and the on-shell formulation of Section~5
 (\drII, \mwII), as a function of \mt, for
 $\mz = 91.187\GeV $ and  $\mh = 60\GeV$. Nonperturbative $t\bar t$
 threshold effects are not included here.}
\end{center}
\end{table}

\begin{table}[h]
\begin{center}
\begin{tabular}{|c|cc|cc|}
 \hline
 \mt    &   \drI      &   \drII     &  \mwI     &  \mwII   \\
 $\mbox{\small[GeV]}$ & \multicolumn{2}{c|}{$\times10^{2}$} &
 \multicolumn{2}{c|}{$\mbox{\small[GeV]}$}  \\
 \hline
 \mz    &      6.461  &      6.462  &   79.862  &   79.862  \\
 100  &      6.199  &      6.200  &   79.911  &   79.911  \\
 110  &      5.907  &      5.908  &   79.965  &   79.964  \\
 120  &      5.612  &      5.613  &   80.018  &   80.018  \\
 130  &      5.309  &      5.309  &   80.072  &   80.072  \\
 140  &      4.994  &      4.994  &   80.128  &   80.128  \\
 150  &      4.665  &      4.665  &   80.186  &   80.186  \\
 160  &      4.320  &      4.320  &   80.246  &   80.246  \\
 170  &      3.959  &      3.957  &   80.308  &   80.308  \\
 180  &      3.579  &      3.577  &   80.372  &   80.372  \\
 190  &      3.179  &      3.176  &   80.439  &   80.439  \\
 200  &      2.760  &      2.756  &   80.508  &   80.508  \\
 210  &      2.319  &      2.313  &   80.579  &   80.580  \\
 220  &      1.855  &      1.848  &   80.653  &   80.654  \\
 230  &      1.369  &      1.360  &   80.730  &   80.731  \\
 240  &      0.858  &      0.846  &   80.809  &   80.810  \\
 250  &      0.322  &      0.307  &   80.890  &   80.892  \\
 \hline
\end{tabular}
\caption{As  Table~VI, for $\mh = 250\GeV$. }
\end{center}
\end{table}

\begin{table}[h]
\begin{center}
\begin{tabular}{|c|cc|cc|}
 \hline
 \mt    &   \drI      &   \drII     &  \mwI     &  \mwII   \\
 $\mbox{\small[GeV]}$ & \multicolumn{2}{c|}{$\times10^{2}$} &
 \multicolumn{2}{c|}{$\mbox{\small[GeV]}$}  \\
\hline
 \mz    &      7.036  &      7.037  &   79.754  &   79.754  \\
 100  &      6.779  &      6.780  &   79.803  &   79.803  \\
 110  &      6.494  &      6.494  &   79.856  &   79.856  \\
 120  &      6.206  &      6.205  &   79.910  &   79.910  \\
 130  &      5.910  &      5.910  &   79.964  &   79.964  \\
 140  &      5.604  &      5.603  &   80.019  &   80.020  \\
 150  &      5.286  &      5.284  &   80.077  &   80.077  \\
 160  &      4.954  &      4.951  &   80.135  &   80.136  \\
 170  &      4.607  &      4.603  &   80.196  &   80.197  \\
 180  &      4.244  &      4.239  &   80.259  &   80.260  \\
 190  &      3.865  &      3.859  &   80.324  &   80.325  \\
 200  &      3.469  &      3.461  &   80.390  &   80.392  \\
 210  &      3.056  &      3.045  &   80.459  &   80.461  \\
 220  &      2.625  &      2.612  &   80.530  &   80.532  \\
 230  &      2.175  &      2.159  &   80.602  &   80.605  \\
 240  &      1.707  &      1.687  &   80.677  &   80.680  \\
 250  &      1.220  &      1.195  &   80.753  &   80.757  \\
 \hline
\end{tabular}
\caption{As  Table~VI, for $\mh = 1000\GeV$. }
\end{center}
\end{table}

\pagebreak
\appendix 

\section*{Appendix A}
\setcounter{section}{1} 
\begin{appendletter} 
In this Appendix we discuss the perturbative corrections of
\Ord{\alphah_s} to the vacuum polarization functions involving quarks.
Defining $\Pi^{V,A}(s,m_1,m_2)$ according to Eqs.~(\ref{E26a},b), we expand
\be
 \Pi^{V,A}(s,m_1,m_2) = \Pi_0^{V,A}(s,m_1,m_2)
    + \frac{\alphah_s}{\pi}\, \Pi_1^{V,A}(s,m_1,m_2) ,
\label{EA1}
\ee
where $s=q^2$ and $m_1$ and $m_2$ are the masses of the virtual quarks
in the loop.
The functions $\Pi_0^{V,A}(s,m_1,m_2)$ have been extensively discussed
in the literature.
They can be gleaned, for example, from Ref.\cite{R40}.
In the \Ord{\alphah_s} terms we consider two limiting cases:
$m_1 = m_2 = m$ (as  occurs in the $\gamma\gamma$, $ZZ$, and
$Z\gamma$ self energies) and $m_1=m, m_2 = 0$ (as applies, to a
very good approximation, to the $(t,b)$ contribution to the $WW$
self energy).

Comparison of Refs.\cite{R22} and~\cite{R25} leads to the following
expressions:
\bea
 \frac{\pi^2}{m^2} \Pi_1^V(s,m,m) &=&
  r \left(-\frac{1}{n-4}-l-4\zeta(3)+\frac{55}{12}\right) + V_1(r) ,
  \label{EA2} \\*
 \frac{\pi^2}{m^2} \Pi_1^A(s,m,m) &=&
  \frac{6}{(n-4)^2} + \frac{2}{n-4} \left(3 l - \frac{r}{2} -
  \frac{11}{4}\right)
  + 3 l^2 - \left(r+\frac{11}{2}\right) l \nonumber\\*
  & &
  + r \left(-4\zeta(3)+\frac{55}{12}\right) + 6 \zeta(3) + 3 \zeta(2) -
  \frac{11}{8} + A_1(r) ,
  \label{EA3} \\*
 \frac{\pi^2}{m^2} \Pi_1^{V,A}(s,m,0) &=&
  \frac{1}{4} \left[
   \frac{6}{(n-4)^2} + \frac{2}{n-4} \left(3 l - \frac{x}{2} -
   \frac{11}{4}\right)
   + 3 l^2 - \left(x+\frac{11}{2}\right) l
  \right. \nonumber\\*
  & &
  \left.
   + x \left(-4\zeta(3)+\frac{55}{12}\right) + 6 \zeta(3) + 3 \zeta(2)
   - \frac{11}{8}\right]
   + F_1(x) ,
  \label{EA4}
\eea
where $r \equiv s/(4 m^2)$, $x \equiv s/m^2$, $l \equiv \ln(m^2/\mu')$
($\mu'$ is the rescaled 't~Hooft mass discussed in Section~2),
$V_1(r)$, $A_1(r)$, and $F_1(x)$ are complicated functions studied in
Ref.\cite{R25}, and the color factor appropriate to $N_c = 3$
has been explicitly included.
We recall that $\zeta(2) = \pi^2/6$ and $\zeta(3) = 1.20206\ldots$\ .
The above expressions can be used to evaluate the \Ord{\alphah_s}
corrections employed in the text.
Thus, the quantity $\piggp{f}(0)$ in Section~2 can be written as
\be
 \piggp{f}(0) = \piggp{l}(0) + \piggp{t}(0) + \Re\piggp{5}(\mzs)
 + \Re\left(\piggp{5}(0)-\piggp{5}(\mzs)\right) ,
\label{EA5}
\ee
where the superscripts $(l)$, $(t)$, and (5) refer to the contributions of the
leptons, the top quark, and the first five quark flavors, respectively.
It is easy to see that the contribution $\piggp{q}(s)$ of quark $q$
to $\pigg(s)$ is
\be
 \piggp{q}(s) = \frac{Q^2_q \Pi^V(s,m_q,m_q)}{s} ,
 \label{EA6}
\ee
where $Q_q$ is the charge of the quark in units of the positron
charge $e$, and $m_q$ is its mass.
In particular,
\be
 \piggp{t}(0) = Q_t^2 (\Pi^V)'(0,\mt,\mt) ,
\label{EA7}
\ee
where the prime on the r.h.s.\  denotes differentiation with respect to $s$.
Using our Eq.~(\ref{EA2}) and Eq.~(18) of Ref.\cite{R25}, we find that the
\Ord{\alphah_s} part of $\piggp{t}(0)$ is $\left(2 \alphah_s/9\pi^3\right)
\left[\ln(\mu'/\mt)\right.$ $\left. + 15/8 - (2(n-4))^{-1}\right]$.
The finite part of this result is included in the second term of
Eq.~(\ref{E7a}), while the pole contributes to the last.
Similarly,
\be
 \Re\piggp{5}(\mzs) = \sum_{q \neq t} Q_q^2
 \frac{\Re\Pi^V(\mzs,m_q,m_q)}{\mzs}.
\label{EA8}
\ee
Using Eq.~(\ref{EA2}), we see that the \Ord{\alphah_s} part of
Eq.~(\ref{EA8}) is
$\left(\alphah_s/4\pi^3\right) \sum_{q \neq t} Q_q^2
 \left[2\ln(\mu'/\mz)\right.$ $\left. + f_2(r_q) - (n-4)^{-1} \right]$,
where $r_q$ and $f_2(r)$ are defined in Section~2.
The finite and pole parts have been included in the third and last terms
of Eq.~(\ref{E7a}), respectively.
As mentioned in Section~2, for
$e^2\Re\left(\piggp{5}(0) - \piggp{5}(\mzs)\right)$ we
employ a recent evaluation~\cite{R10}.
The contributions of \Ord{\alphah}
within the square brackets in the second and third terms of
Eq.~(\ref{E7a}) are simply obtained from the \Ord{\alphah_s} ones by
dividing by the quadratic Casimir coefficient $4/3$ for the fundamental
representation of ${\rm SU(3)_C}$, multiplying by an additional factor of
$Q_q^2$, and replacing $\alphah_s \rightarrow \alphah$.
The contributions of \Ord{\alphah} in $\piggp{l}(0)$ (first term in
Eq.~(\ref{E7a})) can be obtained from those in $\piggp{t}(0)$ by
changing $\mt \rightarrow m_l$ and dividing by $3 Q_t^4$, where the 3 stands
for the color factor.

In the approximation of neglecting the squares of mixing angles, the
quark contribution to $\Awwp{(f)}$ is
\be
 \left[\Awwp{(f)}(s)\right]_{\rm quarks} = - \frac{1}{8} \sum_{\rm doublets}
  \left[\Pi^V(s,m_1,m_2) + \Pi^A(s,m_1,m_2) \right] .
\label{EA9}
\ee
In the \Ord{\alphah_s} part we approximate $m_1 = m_2 = 0$ in the $(u,d)$
and $(c,s)$ contributions and $m_1 = \mt$, $m_2 = 0$ in the $(t,b)$
contribution.
In this limit, $\Pi^A(s,m_1,m_2) = \Pi^V(s,m_1,m_2)$
and we see that the \Ord{\alphah_s} part of Eq.~(\ref{EA8}) is
\be
  \left[\Awwp{(f)}(s)\right]_{\rm QCD} = - \frac{1}{4\pi} \left[ 2
 \alphah_s(\mz)
  \Pi^V_1(s,0,0) + \alphah_s(\mt)\Pi^V_1(s,\mt,0) \right] ,
\label{EA10}
\ee
where the first and the second terms correspond to the contributions of the
light and $(t,b)$ isodoublets, respectively.
The choice of renormalization scale in $\alphah_s$ was explained at the end of
Section~2.
Employing our Eq.~(\ref{EA4}) and Eq.~(20) of Ref.\cite{R25}, we find
\be
  \Re\Pi_1^{V,A}(s,0,0) = - \frac{s}{4\pi^2} \left(
  \frac{1}{n-4} + \ln\frac{s}{{\mu'}^2} + 4 \zeta(3) - \frac{55}{12}
    \right)
\label{EA11}
\ee
and
\bea
& & \Re\left[ \frac{\Awwp{(f)}(\mws) - \Awwp{(f)}(0)}{\mws}\right]_{\rm QCD} =
   \frac{\alphah_s(\mz)}{8 \pi^3} \left(
     \frac{1}{n-4} + \ln\frac{\mws}{{\mu'}^2} + 4 \zeta(3) - \frac{55}{12}
    \right) \nonumber\\*
& &\qquad + \frac{\alphah_s(\mt)}{16 \pi^3} \left[
    \frac{1}{n-4} + \ln\frac{\mws}{{\mu'}^2} + 4 \zeta(3) - \frac{55}{12}
    - 4 \omega_t \left(F_1\left(\frac{1}{\omega_t}\right) - F_1(0)\right)
    + \ln\omega_t \right] .\qquad
\label{EA12}
\eea
Subtracting the pole term, setting $\mu' = \mz$, and multiplying by
$e^2/\scs$, we obtain Eq.~(\ref{E11d}).

Similarly, the quark contribution to $\Azzp{(f)}(s)$ is
\be
 \left[\Azzp{(f)}(s)\right]_{\rm quarks} =
  - \frac{1}{16 \ccs} \sum_{q} \left[
    (1-4\scs C_{3q} Q_q)^2 \Pi^V(s,m_q,m_q) + \Pi^A(s,m_q,m_q)
   \right] ,
\label{EA13}
\ee
where $C_{3q} = +1\, (-1)$ for up (down) members of the doublet and the
sum is over quarks.
In the \Ord{\alphah_s} part we neglect all masses other than \mt\ and
find
\bea
 \left[\Azzp{(f)}(s)\right]_{\rm QCD} &=&
  -\frac{\alphah_s(\mz)}{16\pi\ccs} \sum_{q=1}^4 \left[
   (1-4\scs C_{3q} Q_q)^2 +1
  \right] \Pi^V_1(s,0,0)
  \nonumber\\*
  & &
 - \frac{\alphah_s(\mt)}{16\pi\ccs} \left\{
   \left(1-\frac{8}{3}\scs\right)^2 \Pi^V_1(s,\mt,\mt) + \Pi^A_1(s,\mt,\mt)
   \right.
  \nonumber\\*
  & &  \left.
    + \left[\left(1-\frac{4}{3}\scs\right)^2 + 1\right] \Pi^V_1(s,0,0)
  \right\} ,
\label{EA14}
\eea
where the first and second terms are the contributions of the light and
$(t,b)$ isodoublets, respectively.
Combining Eqs.~(\ref{EA10},\ref{EA14}) and recalling
Eqs.~(\ref{E11d},\ref{EA2}--4), we obtain
\bea
& &
\frac{1}{\mzs} \Re\left[\frac{\Awwp{(f)}(\mws)}{\ccs}
	- \Azzp{(f)}(\mzs) \right]_{\rm QCD}  \nonumber\\*
& &=
 \frac{\alphah_s(\mz)}{8 \pi^3 \ccs} \left[
 c^2\ln{c^2} +
 \left(\frac{1}{n-4} + \ln\frac{\mzs}{{\mu'}^2} + 4 \zeta(3)
 - \frac{55}{12}\right)
 \left(-s^2 + 2 \scs \left(1 - \frac{10}{9} \scs\right)\right)
 \right]
 \nonumber \\*
 & &\phantom{=}+
 \frac{\alphah_s(\mt)}{16\pi^3 \ccs} \left\{
  c^2\ln{c^2} +
  \left(\frac{1}{n-4} + \ln\frac{\mzs}{{\mu'}^2} + 4 \zeta(3)
  - \frac{55}{12}\right)
  \left(-s^2 + 2 \scs \left(1 - \frac{10}{9} \scs\right)\right)
 \right. \nonumber\\*
 & &\phantom{=}\left.
 - c^2 \left(4\omega_t F_1\left(\frac{1}{\omega_t}\right) - \ln\omega_t\right)
 + \left(1 - \frac{8}{3}\scs\right)^2 \left(\mu_t
 V_1\left(\frac{1}{4\mu_t}\right) - {\ln\mu_t\over4}\right)
 + \mu_t A_1\left(\frac{1}{4\mu_t}\right) - {\ln\mu_t\over4}
 \right\}.
\nonumber\\* & & \label{EA15}
\eea
Subtracting again the pole terms, setting $\mu' = \mz$, and multiplying by
$e^2/\scs$, we obtain Eq.~(\ref{E15d}).

Finally, in order to implement the decoupling of the top quark, we need
the quantity $(\alpha/\pi)d$, where $d$ is defined in Eq.~(\ref{E9b}).
{From} Ref.\cite{R40} we find
\be
 \Agz^{(t)}(s)=-{\ehs\over3\hat s\hat c}\left({1\over2}-{4\over3}\scs\right)
\Pi^V(s,\mt,\mt) ,
\label{EA16}
\ee
and, therefore,
\be
 \frac{\alpha}{\pi}\,d =
 -{e^2\over3}\left({1\over2\scs}-{4\over3}\right)
 \left.(\Pi^V)'(0,\mt,\mt)\right|_{\sms}.
\label{EA17}
\ee
The \Ord{\alphah_s} part of Eq.~(\ref{EA17}) is evaluated as in the case of
$\piggp{t}(0)$ (cf.\ discussion after Eq.~(\ref{EA7})).
A relevant combination that occurs in \drcarw\ and \drcar\ is
\be
 e^2 \left.\piggp{t}(0)\right|_{\sms} - \frac{\alpha}{\pi}\,d =
  \frac{e^2}{6 \scs} \left.(\Pi^V)'(0,\mt,\mt)\right|_{\sms} =
  - \frac{\alpha}{\pi}\,\hat{d} ,
\label{EA18}
\ee
where $\hat{d}$ is defined in Eq.~(\ref{E10d}).
\end{appendletter}
\section*{Appendix B}
\begin{appendletter}

In this appendix we outline the derivation of the expressions for
\drcarw\ (cf.\ Eq.~(\ref{E10a})) and \drcar\ (cf.\ Eq.~(\ref{E14})),
corresponding to the strategy in which the decoupling of the top quark
is implemented.

As $\scs(1-\drcarw)$ and $\scs \ccs(1-\drcar)$ are physical observables
(cf.\ Eqs.~(\ref{E2},\ref{E3})), we define \drcarw\ and \drcar\ in such
a way that
\bea
 \left[\scs(1-\drcarw)\right]_{\rm old} &=&
 \left[\scs(1-\drcarw)\right]_{\rm new} ,
 \label{EB1} \\*
 \left[\scs \ccs(1-\drcar)\right]_{\rm old} &=&
 \left[\scs \ccs(1-\drcar)\right]_{\rm new} .
 \label{EB2}
\eea
Here the subscript ``old'' labels the quantities obtained when the \ms\
 counterterms
cancel only the divergent parts involving $\delta$ (cf.\ Eq.~(\ref{E5a})),
as in Ref.\cite{R1}, while new denotes the corrections employed in the present
paper, where the \ms\ counterterms contain small finite parts necessary
to implement the top-quark decoupling in relevant amplitudes.
Our strategy is to retain, in Eqs.~(\ref{EB1}) and (\ref{EB2}), terms of
 \Ord{\alpha^2}
when they involve large logarithmic or $\mts/\mzs$ enhancements, but to
neglect them otherwise.
We also neglect certain corrections of \Ord{\alpha^3}.
Recalling $e_0^2 = \ehs/\hat{Z}_e$ and
Eqs.~(\ref{E7d},\ref{E9a},b,\ref{EA18}), we have
\bea
  \hat s^2_{\rm old} &=& \hat s^2_{\rm new}
  \left( 1 + \frac{\alphah}{\pi}\,d\right),
  \label{EB3} \\*
  \left[\frac{\ehs}{\scs}\right]_{\rm old} &=&
  \left[\frac{\ehs}{\scs}\right]_{\rm new}
  \left( 1 - \frac{\alphah}{\pi}\,\hat{d}\,\right)
  , \label{EB4}
\eea
where $d$ and $\hat{d}$ are defined in Eqs.~(\ref{E9c}) and
(\ref{E10d}), respectively.

The derivation of Eq.~(\ref{E10a}) follows almost immediately.
Inserting Eq.~(\ref{EB3}) into Eq.~(\ref{EB1}), we find
\be
 \left(1 + \frac{\alphah}{\pi}\, d\right)
 \left[1-\drcarw\right]_{\rm old} = \left[1-\drcarw\right]_{\rm new} .
\label{EB5}
\ee
As the only large correction in \drcarw\ is
$(\alpha/\pi)\Delta_{\gamma}$ and $\alphah (1-\alpha
\Delta_{\gamma}/\pi) = \alpha$ (cf.\ Eq.~(\ref{E8c})), we obtain
\be
 (\drcarw)_{\rm new} = (\drcarw)_{\rm old}
 - \frac{\alpha}{\pi}\,d .
\label{EB6}
\ee
The expression for $(\drcarw)_{\rm old}$ in terms of
$\piggp{f}(0)$, $WW$ self energies, and the vertex- and box-diagram
corrections to $\mu$~decay, is given in Eqs.~(7b,8b) of Ref.\cite{R1}.
The only contribution to $(\drcarw)_{\rm old}$ with large
logarithmic enhancement is $e^2\left.\piggp{f}(0)\right.|_{\sms}$ and,
to \Ord{\alpha}, this does not involve \scs.
Therefore, we can replace
$\hat s^2_{\rm old}\rightarrow\hat s^2_{\rm new}$
everywhere in $(\drcarw)_{\rm old}$,
neglecting very small terms of \Ord{\alpha^2}.
Inserting the expression for $(\drcarw)_{\rm old}$ in
Eq.~(\ref{EB6}), we obtain Eq.~(\ref{E10a}).

The derivation of Eq.~(\ref{E14}) is more subtle.
Inserting Eq.~(\ref{EB3}) and
\be
 \hat c^2_{\rm old} = \hat c^2_{\rm new} \left(1 -
\frac{\alphah}{\pi}\left[\frac{\scs}{\ccs}\right]_{\rm new} d\right) ,
\label{EB67}
\ee
into Eq.~(\ref{EB2}), we have
\be
 (\drcar)_{\rm new} = (\drcar)_{\rm old}
 - \frac{\alphah}{\pi}\,d
 \left(1-\left[\frac{\scs}{\ccs}\right]_{\rm new}\right)
 \left[1-\drcar\right]_{\rm old} .
\label{EB7}
\ee
We first consider the second term in Eq.~(\ref{EB7}).
Using Eq.~(\ref{E18b}), this contribution can be written as
$-(\alphah/\pi) d \left(1-\left[\scs/\ccs\right]_{\rm new}\right)
\left[1-\drcarw\right]_{\rm old}c^2/\hat c^2_{\rm old}$,
which, neglecting small \Ord{\alpha^2} terms, becomes $-(\alpha/\pi) d
\left(1-\left[\scs/\ccs\right]_{\rm new}\right)c^2/\hat c^2_{\rm new}$.
Next we analyse the first term on the r.h.s.\ of Eq.~(\ref{EB7}).
According to Eq.~(15b) of Ref.\cite{R1}, it is given by
\be
 (\drcar)_{\rm old} = (\drcarw)_{\rm old} -
  \left[ \frac{\ehs}{\scs\ccs}\left(1-\drcarw\right) \right]_{\rm old}
 \Re\left[ \frac{\Aww(\mws) - \ccs\Azz(\mzs)}{\mzs}\right]_{\sms}^{\rm old} .
\label{EB8}
\ee
Examination of Eq.~(A.8) of Ref.\cite{R1} shows that
\be
\Re\left[ \frac{\Aww(\mws) - \ccs\Azz(\mzs)}{\mzs} \right]^{\rm old}_{\sms} =
  \frac{1}{8 \pi^2} \left(\frac{3 \mts}{8 \mzs} + \frac{1}{2}
  \ln\frac{\mt}{\mz} + \cdots \right) ,
\label{EB9}
\ee
where the ellipses represent non-leading terms (some of which involve
$\ln(\mt/\mz)$ with very small coefficients proportional to $s^2$, \scs,
or $\hat{s}^4$).
The significant point is that the leading contributions in Eq.~(\ref{EB9})
are independent of \scs.
Therefore, in Eq.~(\ref{EB8}) we can replace the last factor in the
second term by an analogous expression with $\hat s^2_{\rm old} \rightarrow
\hat s^2_{\rm new}$, the difference being again small terms of \Ord{\alpha^2}.
Using Eqs.~(\ref{EB4},\ref{EB6},\ref{EB67}), Eq.~(\ref{EB8}) becomes
\bea
 (\drcar)_{\rm old} &=& (\drcarw)_{\rm new}
 + \frac{\alpha}{\pi}\,d
 -\left[\frac{\ehs}{\scs}\right]_{\rm new} \frac{1}{\mzs}
 \left[1-\drcarw-\frac{\alpha}{\pi}\,\hat{d} - \frac{\alpha}{\pi}\,
 d\left(1-\frac{\scs}{\ccs}\right)\right]_{\rm new}
 \nonumber\\*
 & &\qquad\times
 \Re\left[\frac{\Aww(\mws)}{\ccs} - \Azz(\mzs)\right]^{\rm new}_{\sms} .
 \label{EB10}
\eea
Recalling Eqs.~(10,12b,15a) of Ref.\cite{R1}
and neglecting again very small contributions of \Ord{\alpha^2,\alpha^3},
we find that the term proportional to
$(\alpha/\pi) d \left(1-\scs/\ccs\right)$ in Eq.~(\ref{EB10}) can be written as
$(\alpha/\pi) d \left(1-\scs/\ccs\right)\left(c^2/\ccs-1\right)$.
Inserting the above results into Eq.~(\ref{EB7}), we find
\bea
 (\drcar)_{\rm new} &=& (\drcarw)_{\rm new}
 + \frac{\alpha}{\pi}\,d
 \left[1 + \left(1-\frac{\scs}{\ccs}\right)\left(\frac{c^2}{\ccs}-1\right) -
 \left(1-\frac{\scs}{\ccs}\right)\frac{c^2}{\ccs}\right]_{\rm new}
 \nonumber\\*
 & &
  -\left[\frac{\ehs}{\scs}\right]_{\rm new} \frac{1}{\mzs}
    \left(1-(\drcarw)_{\rm new} -
    \frac{\alpha}{\pi}\,\hat{d}\,\right)
    \Re\left[\frac{\Aww(\mws)}{\ccs} - \Azz(\mzs)\right]^{\rm new}_{\sms} .
\qquad\qquad\label{EB11}
\eea
The coefficient of $(\alpha/\pi)d$ simplifies to $\scs/\ccs$ and
Eq.~(\ref{EB11}) equals Eq.~(\ref{E14}).
\end{appendletter}
\section*{Appendix C}
\begin{appendletter}
If terms proportional to squares of mixing angles are neglected,
the fermionic contributions to the $WW$ self energy
become a sum over independent isodoublet contributions and we obtain
\be
 B_0^{(f)} = \frac{\alpha}{12\pi \scs} \sum_{\rm doublets} N_c
 f_1(\omega_+,\omega_-) ,
\label{EC1}
\ee
\bea
 f_1(\omega_+,\omega_-)&=& \ln\left(c^2(\omega_+\omega_-)^{1/2}\right) +
 \left[ 1 - \frac{\omega_+ + \omega_-}{2} - \frac{(\omega_+ -
 \omega_-)^2}{2} \right] 2 \Omega(\omega_+,\omega_-)
 \nonumber\\*
 & & + \frac{3 (\omega_+^2 + \omega_-^2) - (\omega_+ - \omega_-)^4}
 {4(\omega_+-\omega_-)}\ln\frac{\omega_+}{\omega_-}
    - \frac{5}{3}
   + \frac{\omega_+ +\omega_-}{4} +\frac{(\omega_+ -\omega_-)^2}{2} ,
\qquad\qquad\label{EC2}
\eea
where $B_0^{(f)}$ is defined in Eqs.~(\ref{E11a},b), $N_c =3$ and $N_c =1$
for quark and lepton isodoublets, respectively, and $\omega_{\pm} =
m_{\pm}^2/\mws$, with $m_+$ and $m_-$ being the masses of the ``up'' and
``down'' fermions in the isodoublet.
Calling
\be
 C \equiv \frac{\omega_+ +\omega_-}{2} - \frac{(\omega_+ -\omega_-)^2}{4}
 -\frac{1}{4} ,
\label{EC3}
\ee
the function $\Omega(\omega_+,\omega_-)$ is given by
\bea
  \Omega(\omega_+,\omega_-) &=& C^{1/2} \cos^{-1}
    \frac{\omega_+ + \omega_- -1}{2 (\omega_+ \omega_-)^{1/2}}
    \qquad\qquad\qquad\qquad\mbox{for\ } C > 0,
  \label{EC4}\\*
  \Omega(\omega_+,\omega_-) &=& \frac{|C|^{1/2}}{2}
  \ln\left|
    \frac{ \omega_+ + \omega_- -1 - 2|C|^{1/2}}
      {\omega_+ + \omega_- -1 + 2|C|^{1/2}}
  \right|
  \qquad\qquad\mbox{for\ } C < 0.
  \label{EC5}
\eea
Equations~(\ref{EC4},\ref{EC5}) are a simpler version of Eqs.~(A.10,A.11)
of Ref.\cite{R48}.
As expected from the integral representation of the
self energies~\cite{R40}, $f_1(\omega_+,\omega_-)$ is a symmetric
function of $\omega_+$ and $\omega_-$.
As $\omega_- \rightarrow 0$,
\be
  f_1(\omega_+,0) = \ln(c^2 \omega_+) + (\omega_+-1)^2
  \left(1+\frac{\omega_+}{2}\right)\ln\left|\frac{\omega_+ -1}{\omega_+}\right|
  -\frac{5}{3} + \frac{\omega_+}{4} + \frac{\omega_+^2}{2} ,
  \label{EC6}
\ee
while in the limit $\omega_+,\omega_- \rightarrow 0$,
\be
  f_1(0,0) = \ln{c^2} - \frac{5}{3} .
\label{EC7}
\ee
If all the fermion masses other than \mt\ are neglected,
Eqs.~(\ref{EC1},\ref{EC6},\ref{EC7}) lead to Eq.~(\ref{E11c}).
In order to estimate the terms of \Ord{m_b^2/\mws}, we consider the case
$\omega_+ >1$ and $\omega_- \ll 1$.
Including corrections of \Ord{\omega_-}, we have
\be
 f_1(\omega_+,\omega_-) = f_1(\omega_+,0) + \frac{3}{2} \omega_- \left[
 (1+\omega_+^2) \ln\frac{\omega_+}{\omega_+ -1}
 - \omega_+ - \frac{1}{2} \right] + \Ord{\omega_-^2} .
\label{EC8}
\ee
Inserting the second term of Eq.~(\ref{EC8}) in the $(t,b)$-isodoublet
contribution to Eq.~(\ref{EC1}), and identifying $\omega_- = \omega_b =
m_b^2/\mws$ and $\omega_+ = \omega_t = \mts/\mws$, we obtain a very small
shift,
\be
  \left[\delta B_0^{(f)}\right]_{m_b} =
   \frac{3 \alpha}{8 \pi \scs}\,\omega_b \left[
    (1+\omega_t^2) \ln\frac{\omega_t}{\omega_t-1}
    -\omega_t - \frac{1}{2} \right] .
\label{EC9}
\ee
Using $m_b = 4.5 \GeV$ and $\mz = 91.187 \GeV$, we find
that the \Ord{m_b^2/\mws} corrections to Eq.~(\ref{EC1}) are
$\approx 2.5\times10^{-5}$, $5.4\times10^{-6}$, $2.8\times10^{-6}$,
$1.7\times10^{-6}$
for $\mt = \mz$, 150, 200, 250~GeV, respectively.
We note that the cofactor of $\omega_b$
in Eq.~(\ref{EC9}) tends to 0 as $\mt \rightarrow \infty$.
To estimate the contributions of \Ord{m_c^2/\mws}, we set $m_s = 0$ and
keep terms of \Ord{\omega_+} for $\omega_+ << 1$ in Eq.~(\ref{EC6}):
\be
 f_1(\omega_+,0) = f_1(0,0) + \frac{3}{2} \omega_+\left(\ln\omega_+ -
 \frac{1}{2}\right) + \Ord{\omega_+^2} .
\label{EC10}
\ee
Inserting the second term of Eq.~(\ref{EC10}) in the $(c,s)$-isodoublet
contribution to Eq.~(\ref{EC1}), and identifying $\omega_+ =
\omega_c = m_c^2/\mws$, we obtain
\be
 \left[\delta B_0^{(f)}\right]_{m_c} = \frac{3 \alpha}{8 \pi \scs}\,\omega_c
 \left(\ln\omega_c - \frac{1}{2}\right) .
\label{EC11}
\ee
For $m_c = 1.5 \GeV$, this amounts to a shift of $-1.1\times10^{-5}$
to Eq.~(\ref{EC1}).
For the corrections of \Ord{m_{\tau}^2/\mws}, we replace $\omega_c
\rightarrow \omega_{\tau} = m_{\tau}^2/\mws$ and divide by the  color factor 3
in Eq.~(\ref{EC11}).
This leads to a further correction of $-5\times10^{-6}$ to Eq.~(\ref{EC1}).

Turning our attention to $C_0^{(f)}$, defined in Eqs.~(\ref{E15a},b), and
neglecting again the squares of mixing angles, we have
\be
 C_0^{(f)} = \frac{\alpha}{12 \pi \scs \ccs} \left[
  c^2 \sum_{\rm doublets} N_c f_2(\omega_+,\omega_-) -
  \frac{1}{4} \sum_{f} N_c\, g(\mu_f) \right] ,
\label{EC12}
\ee
\bea
 f_2(\omega_+,\omega_-) &=&
 \left[1-\frac{3}{2}(\omega_+ + \omega_-)\right]
 \ln(c^2(\omega_+\omega_-)^{1/2})
 + \left[1-\frac{\omega_+ + \omega_-}{2} - \frac{(\omega_+ -
 \omega_-)^2}{2}\right]
 \nonumber \\*
 & & \times 2 \Omega(\omega_+,\omega_-)
 - \frac{(\omega_+ - \omega_-)^3}{4} \ln\frac{\omega_+}{\omega_-}
 -\frac{5}{3} + \omega_+ + \omega_- + \frac{(\omega_+ - \omega_-)^2}{2},
\qquad
 \label{EC13}\\*
 g(\mu_f) &=& \left[(1-4\scs C_{3f} Q_f)^2 + 1\right]
 \left[\ln\mu_f + \frac{1}{3} + 2 (1 + 2 \mu_f)\left(\Lambda(D_f) -1\right)
 \right]
 \nonumber \\*
 & &
 - 6 \mu_f \left(\ln\mu_f + 2 \Lambda(D_f) - 2\right),
 \label{EC14}
\eea
where the $f$ sum is over quark and lepton flavors, $N_c$ is again the
color factor, $Q_f$ is the electric charge in units of the positron charge
$e$, $C_{3f} = +1\,(-1)$ for up (down) members of the doublet, $\mu_f =
m_f^2/\mzs$, $D_f = 4 \mu_f -1$, and
\bea
  \Lambda(D_f) &=& D_f^{1/2} \tan^{-1}\left(D_f^{-1/2}\right)
   \qquad\qquad\qquad\mbox{for\ }D_f >0 ,
  \label{EC15} \\*
  \Lambda(D_f) &=& \frac{|D_f|^{1/2}}{2} \ln\left|
  \frac{1 + |D_f|^{1/2}}{1 - |D_f|^{1/2}}
    \right|
  \qquad\qquad\mbox{for\ }D_f < 0 .
  \label{EC16}
\eea
As is clear from their structure, the contributions involving $f_2$ and
$g$ in Eq.~(\ref{EC12}) arise from the first and second terms in
Eq.~(\ref{E15a}), respectively.

One readily finds the limiting values
\bea
 f_2(\omega_+,0) &=& \left(1-\frac{3}{2}\omega_+\right) \ln(c^2 \omega_+) +
 (\omega_+ -1)^2 \left(1 + \frac{\omega_+}{2}\right)
   \ln\left|\frac{\omega_+ -1}{\omega_+} \right|
   -\frac{5}{3} + \omega_+ + \frac{\omega_+^2}{2} ,
   \nonumber \\* & & {}  \label{EC17} \\*
 f_2(0,0) &=& \ln{c^2} - \frac{5}{3} ,
   \label{EC18} \\
 g(0) &=&  - \frac{5}{3} \left[ (1 - 4\scs C_{3f} Q_f)^2 + 1\right].
 \label{EC19}
\eea
If the fermion masses other that \mt\ are neglected,
Eqs.~(\ref{EC12},\ref{EC14},\ref{EC17}--19) lead to Eq.~(\ref{E15c}).
To discuss the corrections of \Ord{m_b^2/\mws}, we consider
$f_2(\omega_+,\omega_-)$ for $\omega_+ >1$, $\omega_-\ll 1$ and
$g(\mu_f)$ for $\mu_f \ll 1$.
One readily finds
\bea
 f_2(\omega_+,\omega_-) &=& f_2(\omega_+,0)
  + \frac{3}{2} \omega_- \left[
  (1+\omega_+^2) \ln\frac{\omega_+}{\omega_+-1}
  - \omega_+ - \ln(c^2 \omega_+) \right] + \Ord{\omega_-^2} ,
 \qquad \label{EC20} \\
 g(\mu_f) &=& g(0) - 48 \mu_f \scs Q_f( 2 \scs Q_f - C_{3f}) +
   \Ord{\mu_f^2}.
  \label{EC21}
\eea
Identifying $\omega_- = \omega_b$, $\omega_+ = \omega_t$,
$\mu_f = \mu_b = m_b^2/\mzs$ and inserting Eqs.~(\ref{EC20},\ref{EC21})
into Eq.~(\ref{EC12}), we find for the leading correction of \Ord{\mu_b}
\be
 \left[\delta C_0^{(f)}\right]_{m_b} = \frac{3 \alpha}{8 \pi \scs \ccs}
 \,\mu_b
 \left[ (1+\omega_t^2) \ln\frac{\omega_t}{\omega_t -1}
  - \omega_t - \ln(c^2 \omega_t) - \frac{8}{3} \scs \left(1 - \frac{2}{3}
 \scs\right)\right] .
\label{EC22}
\ee
Using the same input values as in Eq.~(\ref{EC9}), we find that
the \Ord{m_b^2/\mws} corrections to Eq.~(\ref{EC12}) are
$\approx  2.5\times10^{-5}$, $-6.7\times10^{-6}$, $-1.6\times10^{-5}$,
$-2.3\times10^{-5}$ for $\mt = \mz$, 150, 200, 250~GeV, respectively.
To estimate the corrections of \Ord{m_c^2/\mws}, we set $\omega_- = 0$
and keep terms of \Ord{\omega_+} for $\omega_+ \ll 1$ in Eq.~(\ref{EC17}).
In conjunction with Eq.~(\ref{EC21}), this gives
\be
 \left[\delta C_0^{(f)}\right]_{m_c} = \frac{3 \alpha}{8 \pi \scs\ccs}
 \,\mu_c
  \left[
   \frac{16}{3} \scs \left(\frac{4}{3} \scs -1\right) - \ln{c^2}
  \right] ,
\label{EC23}
\ee
where $\mu_c = m_c^2/\mzs$.
For $m_c = 1.5 \GeV$, this is $\approx -8\times10^{-7}$.
The corrections of \Ord{m_{\tau}^2/\mws} are even smaller on account of
the absence of the color factor.

There are also very small corrections of \Ord{m_b^2/\mzs} and
\Ord{m_c^2/\mzs} associated with the $f_1(r_q)$ contribution to $e^2
\piggp{f}(0)$ (cf.\ Eq.~(\ref{E7a})).
Their magnitude is $\approx +5\times10^{-6}$.

Putting all these small \Ord{m_f^2/\mws} corrections together, we see that for
$\mt = \mz$, 150, 200, 250~GeV they amount to
(1.4, $-0.6$, $-0.8$, $-0.9)\times10^{-5}$ in the case of \drcarw\ and
$(-1.0$, $+0.2$, $+0.9$, $+1.5)\times10^{-5}$ in the case of \drcar.
On the other hand, $\dr = \drcarw - (\ccs/s^2)(\drcarw - \drcar)$
and the small corrections are enhanced in the second term, leading to
shifts of $(-6.5$, $+2.1$, $+5.1$, $+7.8)\times10^{-5}$ for the same
values of \mt.
\end{appendletter}



\begin{thebibliography}{100}

\bibitem{R1}
G.~Degrassi, S.~Fanchiotti, and A.~Sirlin,
\newblock Nucl.\ Phys.\ {\bf B351}, 49 (1991).

\bibitem{R2}
A.~Sirlin,
\newblock Phys.\ Rev.\ {\bf D22}, 971 (1980).

\bibitem{R3}
A.~Sirlin,
\newblock Phys.\ Rev.\ {\bf D29}, 89 (1984).

\bibitem{R4}
M.~Consoli, W.~Hollik, and F.~Jegerlehner,
\newblock Phys.\ Lett.\ {\bf B227}, 167 (1989).

\bibitem{R5}
G.~Burgers, F.~Jegerlehner, B.~Kniehl, and J.~K\"uhn,
\newblock in {\em $Z$ Physics at LEP~1}, ed.\ by G.~Altarelli, R.~Kleiss,
and C.~Verzegnassi,  Report CERN~89-08, Geneva, (1989) Vol.~1, p.~55.

\bibitem{R6}
J.~J. van der Bij and F.~Hoogeveen,
\newblock Nucl.\ Phys.\ {\bf B283}, 477 (1987).

\bibitem{R7}
R. Barbieri, M.~Beccaria, P.~Ciafaloni, G.~Curci, and A.~Vicer\'e,
\newblock Phys.\ Lett.\ {\bf B288}, 95 (1992);
M.~Beccaria, private communication.

\bibitem{R8}
S.~Fanchiotti and A.~Sirlin,
\newblock in {\em B\'{e}g Memorial Volume}, ed.\ by A.~Ali and
  P.~Hoodbhoy (World Scientific, Singapore, 1991), p.~58.

\bibitem{R9}
H.~Burkhardt, F.~Jegerlehner, G.~Penso, and C.~Verzegnassi,
\newblock Z.\ Phys.\ {\bf C43}, 497 (1989).

\bibitem{R10}
F.~Jegerlehner,
\newblock in {\em Proceedings of the 1990 Theoretical Advanced Study
Institute in Elementary Particle Physics}, ed.\ by
P.~Langacker and M.~Cveti\v{c} (World Scientific, Singapore, 1991) p.~476.

\bibitem{R11}
A.~Sirlin,
\newblock Phys.\ Lett.\ {\bf B232}, 123 (1989).

\bibitem{R12}
S.~Fanchiotti and A.~Sirlin,
\newblock Phys.\ Rev.\ {\bf D41}, 319 (1990).

\bibitem{R13}
A.~Sirlin,
\newblock Phys.\ Rev.\ Lett.\ {\bf 67}, 2127 (1991);
\newblock Phys.\ Lett.\ {\bf B267}, 240 (1991).

\bibitem{R14}
M.~Consoli and A.~Sirlin,
\newblock in {\em Physics at LEP}, ed.\ by J.~Ellis and R.~Peccei,
\newblock Report CERN~86-02, (1986) Vol.~1, p.~63.

\bibitem{R15}
S.~Willenbrock and G.~Valencia,
\newblock Phys.\ Lett.\ {\bf B259}, 373 (1991);
R.~G.~Stuart,
\newblock Phys.\ Lett.\ {\bf B262}, 113 (1991);
{\it ibid.} {\bf B272}, 353 (1991);
T.~Bhattacharya and S.~Willenbrock,
\newblock Brookhaven National Laboratory Report BNL-48937.

\bibitem{R16}
R.~Tanaka,
\newblock talk given at the {\em XXVI Int.\ Conf.\ on High Energy Physics},
Dallas, Texas, August~1992.

\bibitem{R17}
G.~Degrassi and A.~Sirlin,
\newblock Nucl.\ Phys.\ {\bf B352}, 342 (1991).

\bibitem{R18}
U. Amaldi et al.,
\newblock Phys.\ Rev.\ {\bf D36}, 1385 (1987);
G.~Costa, J.~Ellis, G.~L.~Fogli, D.~V.~Nanopoulos, and F.~Zwirner,
\newblock Nucl.\ Phys.\ {\bf B297}, 244 (1988).

\bibitem{R19}
P.~Langacker, M.~Luo, and A.~Mann,
\newblock Rev.\ Mod.\ Phys.\ {\bf 64}, 87 (1992).

\bibitem{R20}
G.~Altarelli, R.~Barbieri, and S.~Jadach,
\newblock Nucl.\ Phys.\ {\bf B369}, 3 (1992).

\bibitem{R21}
P.~Langacker,
\newblock Conference Summary, in {\em XXVII\/${}^{th}$ Rencontres de Moriond on
Electroweak Interactions and Unified Theories}, Les Arcs, France, March
1992,
\newblock University of Pensylvania Report UPR-0512T.

\bibitem{R22}
A.~Djouadi and C.~Verzegnassi,
\newblock Phys.\ Lett.\ {\bf B195}, 265 (1987);
A.~Djouadi,
\newblock Nuovo Cim.\ {\bf 100A}, 357 (1988).

\bibitem{R23}
T.~H.~Chang, K.~J.~F.~Gaemers, and W.~L.~van~Neerven,
\newblock Nucl.\ Phys.\ {\bf B202}, 407 (1982).

\bibitem{R24}
B.~A.~Kniehl, J.~H.~K\"uhn, and R.~G.~Stuart,
\newblock Phys.\ Lett.\ {\bf B214}, 621 (1988);
\newblock also in {\em Polarization at LEP}, ed.\ by G.~Alexander et al.,
Report CERN~88-06, (1988) Vol.~1, p.~158;
B.~A.~Kniehl,
\newblock Comput.\ Phys.\ Commun.\ {\bf 58}, 293 (1990).

\bibitem{R25}
B.~A.~Kniehl,
\newblock Nucl.\ Phys.\ {\bf B347}, 86 (1990).

\bibitem{R26}
F.~Halzen and B.~A.~Kniehl,
\newblock Nucl.\ Phys.\ {\bf B353}, 567 (1991).

\bibitem{R27}
V.~S.~Fadin and V.~A.~Khoze,
\newblock JETP Lett.\ {\bf 46}, 525 (1987);
\newblock Sov.\ J.\ Nucl.\ Phys.\ {\bf 48}, 309 (1988).

\bibitem{R28}
M.~J.~Strassler and M.~E.~Peskin,
\newblock Phys.\ Rev.\ {\bf D43}, 1500 (1991).

\bibitem{R29}
W.~Kwong,
\newblock Phys.\ Rev.\ {\bf D43}, 1488 (1991).

\bibitem{R30}
B.~A.~Kniehl and A.~Sirlin,
\newblock Nucl.\ Phys.\ {\bf B371}, 141 (1992).

\bibitem{R31}
B.~A.~Kniehl and A.~Sirlin,
\newblock  preprint DESY~92-102 and NYU-TR~92/09/04 (September 1992),
Phys.\ Rev.\ D (to appear).

\bibitem{R32}
W.~J.~Marciano and J.~L.~Rosner,
\newblock Phys.\ Rev.\ Lett.\ {\bf 65}, 2963 (1990).

\bibitem{R33}
W.~J.~Marciano,
\newblock Annu.\ Rev.\ Nucl.\ Part.\ Sci.\ {\bf41} (1991) 469.

\bibitem{R34}
W.~A.~Bardeen, A.~J.~Buras, D.~W.~Duke, and T.~Muta,
\newblock Phys.\ Rev.\ {\bf D18}, 3998 (1978);
A.~J.~Buras,
\newblock Rev.\ Mod.\ Phys.\ {\bf 52}, 199 (1980).

\bibitem{R35}
W.~J.~Marciano, private communication.

\bibitem{R36}
F.~Abe et al.\ (CDF Collaboration),
\newblock Phys.\ Rev.\ Lett.\ {\bf68}, 447 (1992);
\newblock Phys.\ Rev.\ {\bf D45}, 3921 (1992).

\bibitem{R37}
F.~Abe et al.\ (CDF Collaboration),
\newblock Phys.\ Rev.\ Lett.\ {\bf65}, 2243 (1990);
\newblock Phys.\ Rev.\ {\bf D43}, 2070 (1991);
J.~Alitti et al.\ (UA2 Collaboration),
\newblock Phys.\ Lett.\ {\bf B276}, 354 (1992).

\bibitem{R50}
K.~Hikasa et al.\ (Particle Data Group),
\newblock Phys.\ Rev.\ {\bf D45}, S1 (1992), Part II.

\bibitem{R38}
S.~Bethke,
\newblock J.\ Phys.\ {\bf G17}, 1455 (1991).

\bibitem{R39}
S.~Sarantakos, A.~Sirlin, and W.~J.~Marciano,
\newblock Nucl.\ Phys.\ {\bf B217}, 84 (1983).

\bibitem{R40}
W.~J.~Marciano and A.~Sirlin,
\newblock Phys.\ Rev.\ {\bf D22}, 2695 (1980).

\bibitem{R41}
V.~A.~Novikov, L.~B.~Okun, M.~A.~Shifman, A.~I.~Vainshtein, M.~B.~Voloshin,
and V.~I.~Zakharov,
\newblock Phys.\ Rep.\ {\bf 41C}, 1 (1978);
M.~A.~Shifman, A.~I.~Vainshtein, and V.~I.~Zakharov,
\newblock Nucl.\ Phys.\ {\bf B147}, 385 (1979).

\bibitem{R42}
B.~Grinstein and M.-Y.~Wang,
\newblock Nucl.\ Phys.\ {\bf B377}, 480 (1992).

\bibitem{R43}
J.~Gasser and H.~Leutwyler,
\newblock Phys.\ Rep.\ {\bf 87}, 77 (1982).

\bibitem{R44}
F.~Jegerlehner,
\newblock in {\em Progress in Particle and Nuclear Physics}, Vol.~27,
ed.\ A.~Faessler (Pergamon Press, Oxford, 1991), p.~1;
A.~Sirlin (unpublished).

\bibitem{R45}
H.~Georgi and H.~D.~Politzer, Phys.\ Rev.\ {\bf D14}, 1829 (1976).

\bibitem{R46}
M.~Veltman,
\newblock Nucl.\ Phys.\ {\bf B123}, 89 (1977);
M.~S.~Chanowitz, M.~A.~Furman, and I.~Hinchliffe,
\newblock Phys.\ Lett.\ {\bf 78B}, 285 (1978).

\bibitem{R47}
M.~Consoli, S.~Lo~Presti, and L.~Maiani,
\newblock Nucl.\ Phys.\ {\bf B223}, 474 (1983);
F.~Halzen, Z.~Hioki, and M.~Konuma,
\newblock Phys.\ Lett.\ {\bf 126B}, 129 (1983);
Z.~Hioki,
\newblock Nucl.\ Phys.\ {\bf B229}, 284 (1983);
\newblock Phys.\ Rev.\ {\bf D45}, 1814 (1992);
\newblock Mod.\ Phys.\ Lett.\ {\bf A7}, 1009 (1992);
\newblock Tokushima Report 92-02.

\bibitem{R48}
S.~Bertolini and A.~Sirlin,
\newblock Nucl.\ Phys.\ {\bf B248}, 589 (1984).

\bibitem{R49}
A.~Sirlin,
\newblock Higher-order electroweak corrections,
\newblock lecture presented at the {\em Workshop on Precise Electroweak
Measurements}, UC-Santa Barbara, California, February~1991.

\bibitem{R51}
L.~Rolandi,
\newblock talk given at the {\em XXVI Int.\ Conf.\ on High Energy Physics},
Dallas, Texas, August~1992.

\end{thebibliography}
\end{document}